\documentstyle[epsf]{mn}
\title[Disc--corona energtics in the VHS]
{Disc--corona energetics in the 
Very High State of Galactic Black Holes}

\author[C. Done, A. Kubota]
{Chris Done$^1$ and Aya Kubota$^2$\\
$^1$Department of Physics, University of Durham, South Road, Durham 
DH1 3LE, UK; chris.done@durham.ac.uk\\
$^2$Cosmic Radiation Laboratory, Institute of Physical and Chemical Research,\\
2-1 Hirosawa, Wako-shi, Saitama, 351-0198, Japan; aya@crab.riken.jp\\ 
}

\date{Accepted 2005 * **.
      Received 2005 * **;
      in original form 2005 * **}

\begin{document}

\maketitle

\begin{abstract}

The X-ray spectra of Galactic binary systems dominated by a
quasi-thermal component (disc dominated or high/soft state) are well
described by a standard Shakura-Sunyaev disc structure down to the last
stable orbit around the black hole. This is not the case in the very
high (or steep power law) state, where the X-ray spectra show both a
strong disc component and strong, steep tail to higher energies. 
We use simultaneous optical-ASCA-RXTE data from the black hole transient
XTE~J$1550-564$ as a specific example of two such spectra, where the
power emitted in the tail is more than 50 per cent of the bolometric
luminosity.  These have disc spectra which are significantly lower in
temperature than those seen from the same source at the same luminosity
in the high/soft state. If these give a true picture of the disc then
either the disc emissivity has reduced, and/or the disc truncates above
the last stable orbit. 

However, it is often assumed that the tail is produced by Compton
scattering, in which case its shape in these spectra requires that the
Comptonising region is marginally optically thick ($\tau\sim$~2--3), and
covers a large fraction of the inner disc. This will distort our view of
the disc, especially of the hottest temperature material. We build a
theoretical model of a Comptonising corona over an inner disc, and fit
this to the data, but find that it still requires a large increase in
inner disc radius for a standard disc emissivity. Instead it seems more
probable that the disc emissivity {\em changes} in the presence of the
corona. We implement the specific inner disc-corona coupling model of Svensson
\& Zdziarski (1994), in which some fraction $f$ of the accretion power
is dissipated in the corona leaving only a fraction $1-f$ to be
dissipated in the optically thick disc.  We show that this can explain
the low temperature/high luminosity disc emission seen in the very high
state with only a small increase in radius of the disc. While this
inferred disc truncation is probably not significant given the model
uncertainties, it is consistent with the low frequency QPO and gives
continuity of properties with the low/hard state spectra. 

\end{abstract}


\section{Introduction}

Galactic black hole binaries at high mass accretion rates show spectra which are 
often dominated by a soft, quasi-thermal component (e.g. the review by Tanaka \& 
Lewin 1995). This shape matches very well with that expected from an optically 
thick, geometrically thin accretion disc, as described in the seminal paper by 
Shakura \& Sunyaev (1973). This identification is strengthened by observations 
of variability in which the temperature, $T_{in}$, and luminosity, $L$, of this 
quasi-thermal component change together in such a way as to indicate that the 
size of the emitting structure remains approximately constant ($L\propto 
T_{in}^4$). The last stable orbit is the only obvious candidate for such a 
radius, so this gives compelling evidence for a key prediction of general 
relativity, as well as providing an observable diagnostic of the mass and spin 
of the black hole (Ebisawa et al 1994; Kubota et al 2001; Gierli{\'n}ski \& Done 
2004; Kubota \& Done 2004; hereafter KD04). The most recent models of accretion 
disc spectra are able to reproduce this behavior (Davis et al 2005; Davis, Done 
\& Blaes 2006). These supersede earlier models which showed strong deviations 
from the $L\propto T_{in}^4$ relation, but which did not include metal opacity 
nor treat the difference in vertical structure between the radiation and gas 
pressure dominated regimes (Merloni, Fabian \& Ross 2000).

The disc dominated spectra always contain some small fraction of emission in a 
tail extending to much higher energies. However, some high luminosity spectra 
have a rather different spectral shape, where the tail contains a much larger 
fraction of the bolometric luminosity. The disc emission in these spectra 
(variously termed very high or steep power law dominant state, hereafter VHS; 
Miyamoto et al. 1991; McClintock \& Remillard 2003) is less obvious due to the 
lower contrast against the strong continuum. Nonetheless, it can be identified 
in spectral fitting, but its derived parameters (temperature and luminosity) are 
somewhat dependent on how the tail is modelled, which in turn depends on the 
radiation processes involved. A power law or broken power law description for 
the tail might be appropriate if the emission is from nonthermal electrons in 
the jet, but these models require fine tuning in order to match all the 
observational constraints. Conversely, Comptonisation models, where the disc is 
the source of seed photons, rather naturally reproduce both multiwavelength 
spectral constraints and X-ray variability patterns (see \S 3), so here
we explore the consequences of such models. 

Firstly, the disc temperature in GBH is close to (or within!) the observed 
X-ray bandpass, so the low energy break in the Comptonised component is predicted to 
be present, giving a rather different spectrum at low energies to that of a power law
which extends unbroken through the disc 
emission. Thus using proper Comptonisation models can significantly change the 
parameters of the inferred disc component (Done, $\dot{\rm Z}$ycki \& Smith 
2002) and distort the measured luminosity-temperature relation (Kubota et al 
2001; Kubota \& Makishima 2004; KD04). This helps considerably in interpreting 
data from the weak VHS, where the disc contributes more than half of the total 
bolometric luminosity. A power law model for the tail in these data gives 
derived disc temperatures which are significantly higher than those seen from 
the same source at the same disc luminosity in the high/soft state. Using proper 
Comptonisation models instead of the power law approximately recovers the 
$L\propto T_{in}^4$ behavior (Kubota et al 2001; Kubota \& Makishima 2004; 
KD04). The match to the high/soft state data is even better with an additional 
correction to the observed disc luminosity to account for the photons which are 
scattered into the Comptonised spectrum (Kubota \& Makishima 2004; KD04). 

However, the most strongly Comptonised VHS (where the disc is less than half of 
the bolometric luminosity) have disk spectra which show significantly lower 
temperatures than those seen at the same disc luminosity in the high/soft state. 
These are inconsistent with a constant disc radius even with proper 
Comptonisation models and correction for scattering (KD04). This could indicate 
that the disc is truncated above the last stable orbit (KD04), but here we 
explore whether this conclusion is required with a more sophisticated approach 
to Comptonisation. The parameters derived for the hot electrons indicate that 
the Compton cloud is marginally optically thick, with $\tau\sim$~2--3. This 
material plainly must intercept the disc photons in order to Comptonise them, so 
will also distort our view of the disc (KD04). An optically thick corona over 
the inner disc could simply hide the hottest part of the disc from view.
More likely, the power dissipated in the corona can {\em reduce} the disc emissivity (e.g. 
Svensson \& Zdziarski 1994) since the energy in the corona must derive ultimately from 
the accretion flow. We build explicit models of an inner disc- corona 
to explore these ideas, and fit them to two strong VHS spectra 
(simultaneous 0.7-- 200~keV ASCA and RXTE data) of the Galactic black hole 
binary XTE~J$1550-564$ to constrain the possible geometries of the disk-corona 
in this state.

In \S2, we review the spectral parameters (luminosity and temperature) of the 
geometrically thin disk in the disk dominated high/soft state. In \S3 we 
describe the simultaneous ASCA-RXTE observations of the broadband spectra. We 
compare the derived disc emission in the VHS to that seen in the high/soft 
state, showing that it implies a very different disc area and/or emissivity 
(\S4). In \S5 we describe our model of an inner disc corona above a standard 
emissivity disc. Fitting this to the data shows explicitly that optical depth 
effects alone cannot hide the inner disc, and that this model requires a large 
change in disc area compared to the high/soft state. We modify this in \S6 to 
incorporate energetic coupling of the inner disc and corona as in Svensson \& 
Zdziarski (1994). The reduction in disc emissivity and temperature under the 
corona means this model can fit the observed spectra with only a small - or 
possibly no - increase in disc area. 

We review all the results of the models in \S7, and then incorporate 
additional constraints from other spectral and timing characteristics 
of this state in \S8. We show 
that an optically thick Comptonisation in an inner disc corona is not 
inconsistent with the presence of high frequency QPO's.  However, it does 
conflict with claims of a strong, extremely broad iron line in these spectra
(Miller et al 2004). We show that there is strong evidence supporting 
Comptonisation as the emission mechanism for the tail (\S9), whereas
there is an alternative explanation for the iron line, in which its width
is overestimated due to the unrecognized presence of absorption lines
(Done \& Gierlin\'ski 2006). 

All the Comptonisation models considered here {\em require} 
that the inner disc changes in extent and/or emissivity in response to the 
presence of the corona. There are no solutions in which the inner disc has the 
same radial structure as that seen in the high/soft state. 
 
\section{The disk dominant high/soft state as a template for the VHS geometry}

In all the following we use the  
binary system parameters determined by the 
optical studies of Orosz et al. (2002) i.e. mass of 8.4--11.2 $M_\odot$, 
distance, $D$, of $\sim$~5~kpc and binary inclination angle, $i$, of $70^\circ$ 
(Orosz et al. ~2002). However, 
we are primarily interested in {\em changes} in the disc emitting area rather 
than absolute values i.e. in comparing the disc structure in the VHS with that 
seen in the disc dominated high/soft states. The high/soft states in XTE~J$1550-
564$ have been well fit using the {\sc diskbb} model (Mitsuda et al. 1984) in 
XSPEC (Kubota \& Makishima~2004; Gierli\'nksi \& Done 2004; KD04), giving a clear 
$L\propto T_{in}^4$ relationship, implying constant area, over a factor 100 
change in disc luminosity. Such an obviously thermal relationship between 
luminosity and temperature from {\sc diskbb} fits is now predicted by the best 
current models of the disc spectra which include radiative transfer with full 
metal opacities and self consistent vertical structure as well as relativistic 
effects (Davies et al 2005). 

The apparent inner disc radius of this source, $r_{\rm in}$, is obtained from 
the normalization of {\sc diskbb} model, defined as $r_{\rm in}^2\cos 
i/(D/10{\rm kpc})^2$. The average value inferred from all the high/soft state 
constant area {\sc diskbb} results is $59$ km. 
However, the 'true' inner disc radius is related to this 
apparent radius via a series of correction factors (for the stress free inner 
boundary condition, effective temperature not equal to colour temperature and 
relativistic effects; e.g. Kubota et al 1998; 2001).  However, the fractional 
{\em change} in inferred disc radius is unaffected by uncertainties in distance, 
inclination and correction factors, so we give the apparent inner radius, 
$r_{\rm in}$ as a {\em ratio} of {\sc diskbb} normalisation seen in these fits 
to that derived from all the high/soft state data in Kubota \& Makishima (2004). 
These ratios are not sensitive to any of the binary system uncertainties, so can 
sensitively show the {\em change} in the disc properties between the high/soft 
and VHS. 

\section{ASCA-{\it RXTE} simultaneous observations}

ASCA observations of this source were performed three times, on 1998 September 
12, 23, and 1999 March 17.  There are simultaneous {\it RXTE} observations 
corresponding to each ASCA observation. The second data set shows a strongly 
Comptonised VHS spectrum, and was included in the analysis of KD04. Here we also 
study the first simultaneous data set, which is at the boundary between the 
low/hard state and the strongly Comptonised VHS. We do not use the third ASCA 
dataset as the gain was changed on the corresponding {\it RXTE} data dataset 
midway through the observation, and the spectrum is only a weakly comptonised 
VHS.  The observational ID's of the corresponding RXTE data are 30188-06-05-00 
and 30191-01-10-00, for the first and second ASCA observations, respectively. We 
use the data as extracted by KD04 (standard procedures for selecting good PCA 
and HEXTE data, with PCA responses derived for each observation using {\it 
pcarsp} v8.0: see Appendix for a comparison with {\it 
pcarsp} v10.1). We use the PCA data from 3--20~keV, and HEXTE from 20-200~keV. 

The ASCA GIS events were extracted from circular regions of $6'$
radius centered on the image peak, after selecting good time intervals
in a standard procedure.  Net exposures of 1.4~ks and 2.4~ks were
obtained for the simultaneous pointing on the first and second
observations, respectively. The dead time correction fractions for the
first ASCA data set are 76.1\% for GIS2 and 77.9\% for GIS3, determined
by reference to the count-rate monitor data with little dead time,
using the method described in Makishima et al (1996).  Similarly, for
the second observation, these are calculated as 85.6\% and 87.3\% for
GIS2 and GIS3, respectively.  We only use the GIS data (0.7--10~keV) 
since the SIS is strongly affected by pileup. 

In order to take into account the calibration uncertainties, we add 1\% 
systematic errors to the GIS spectra (Makishima et al 1996) and 0.5\% systematic 
errors to the PCA. We also include 0.5\% systematic uncertainty on the HEXTE
data to be conservative, though this does not significantly change the fit
(see Appendix). We allow $N_H$ to be free (described 
using {\sc wabs}) as the low energy bandpass of the ASCA GIS constrains this 
parameter. The data from different instruments are fitted with same model 
parameters except that relative normalization factors are free. Hereafter, we 
use the normalization of the PCA in the fitting results so as to be able to 
directly compare with the PCA analysis of Kubota \& Makishima (2004) for the 
disc spectra. Using this we find the observed (i.e. absorbed) 0.7-- 200~keV 
fluxes of $7.2\times10^{-8}~{\rm erg~s^{-1}~cm^{-2}}$ and $8.3\times10^{-8}~{\rm 
erg~s^{-1}~cm^{-2}}$ for the first and second observations, respectively, 
corresponding to observed luminosities of $2.2\times10^{38}~{\rm erg~s^{-1}}$ 
and $2.5\times10^{38}~{\rm erg~s^{-1}}$ for isotropic emission. Correcting for 
absorption and the energy band gives a bolometric luminosity of $\sim 4.0$ and 
$4.7\times10^{38}~{\rm erg~s^{-1}}$ i.e. about 35 per cent of the Eddington 
limit.

%sim1
%7.2037E-08 ergs)cm**-2 s**-1 (  0.700-200.000)
%Nh=0 ( 9.2828E-08 ergs)cm**-2 s**-1 (  0.700-200.000)
%sim2
%8.3928E-08 ergs)cm**-2 s**-1 (  0.700-200.000)
%Nh=0 1.1115E-07 ergs)cm**-2 s**-1 (  0.700-200.000)

\section{Continuous disc-corona}

\subsection{Fitting the data}

\begin{figure*}
\begin{center}
\leavevmode
\epsfxsize=0.65\textwidth \epsfbox{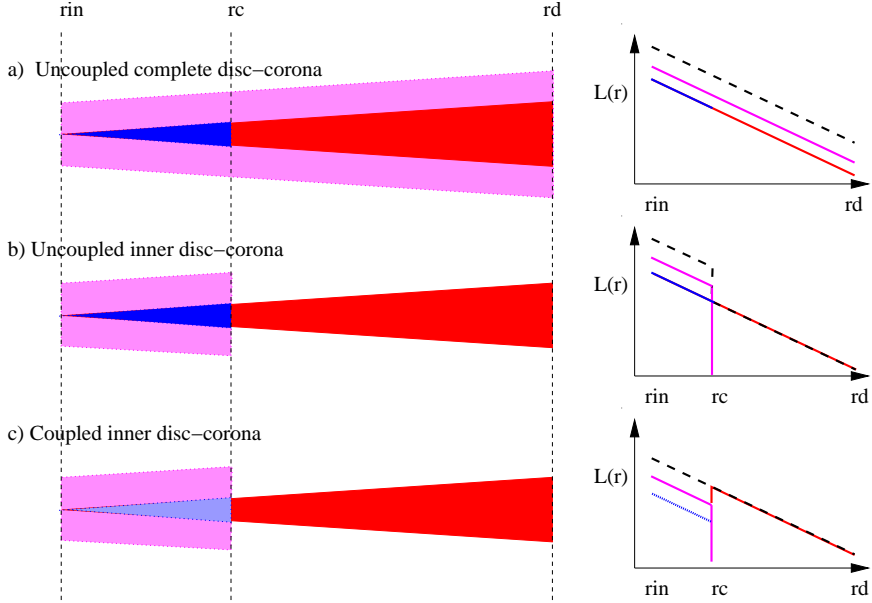}
\end{center}
\caption{The schematic 
geometry envisaged for each of the models described in the text.
To the right of each geometry is a sketch of the associated emissivity as a 
function of radius for the outer disc (red), inner disc (blue),
corona (magenta) and total (black)
accretion flow.}
\label{fig:sketch1}

\end{figure*}

These VHS spectra are dominated by a strong, steep continuum rather than the 
disc, though the disc can still be seen. The continuum shape is complex: fits 
using a single power law (together with the disc, and including a smeared edge 
and Gaussian to phenomenologically model the reflection components) are 
completely statistically unacceptable, with $\chi^2_\nu$ of $2538/195$ and 
$1718/195$, for the first and second datasets, respectively. This was also shown 
by Sobczak et al (1999), who had to fit the PCA and HEXTE data separately in 
order to get acceptable fits with a power law continuum. However, the continuum 
curvature is complex: a single thermal Comptonisation model for the tail (with 
disc, smeared edge and Gaussian line) gives a much improved fit over a power 
law, but the fits are not formally acceptable at $\chi^2_\nu$ of $402.9/193$ 
and $261.5/193$, respectively. 

The complex form of the tail is shown in more detail by Gierli\'nski \& Done 
(2003), who used OSSE data together with the PCA and HEXTE to extend the 
bandpass beyond 500~keV for times around that of the second dataset used here. 
These clearly showed that the curvature seen in the tail cannot be matched by 
Comptonisation from thermal or non-thermal electrons (see Fig. 5 of
Gierli\'nski \& Done 2003). Instead the continuum 
shows features of {\em both} thermal {\em and} non-thermal electrons (see also 
Zdziarski et al 2001 for similar spectra from GRS~1915+105). Here we follow KD04 
and approximate this by using the {\sc thcomp} model (Zdziarski, Johnson \& 
Magdziarz 1996; now publically available as a local model in
XSPEC\footnote{http://heasarc.gsfc.nasa.gov/docs/xanadu/xspec/newmodels.html})
to describe the thermal Compton scattering, together with a weak 
power law of $\Gamma$ fixed at 2 to approximate the higher energy (non- thermal) 
emission. We have checked that our results on the disc temperature and 
luminosity are not dependent on this approximation for the shape of the high 
energy spectrum by using the full hybrid thermal-nonthermal Comptonisation code 
{\sc eqpair} (Coppi 1999) on our final model of the Comptonised disc-corona 
described in \S6.

The seed photons in {\sc thcomp} are set to be from the disc (giving a slightly 
different low energy rollover to that produced by the {\sc comptt} model in the 
XSPEC general release which assumes a Wien distribution). This is equivalent to 
assuming that the Comptonising corona covers the whole disc, as shown 
schematically in Fig.~\ref{fig:sketch1}a.

Following KD04 we use several different descriptions of the reflected emission, 
to try to assess the effect on the derived disc parameters of systematic 
uncertainties in current reflection models. Firstly we use the Gaussian line and 
smeared edge to give a phenomenological description of reflection (SEG).
Then we replace these components with the physical reflection model 
incorporated in {\sc thcomp} (REF) which includes both full reflected 
continuum and self--consistent iron line emission for arbitrary ionization state 
of the reflected material ($\dot{\rm Z}$ycki, Done \& Smith 1999). 
Ionization is important as the lack of photoelectric opacity in 
ionized material means reflection contributes to the low
energy spectrum so can affect the derived disc temperature and
luminosity.

However, the ionised reflection in {\sc thcomp} 
does not include Compton up and downscattering of the reflected emission in 
the disc itself. This is important for highly ionized and/or hot discs, 
and leads to strong distortions on the predicted line and edge shape in the 
reflected emission which can affect the spectral fitting, either by distorting 
the inferred solid angle of reflecting material, or its ionization or amount of 
relativistic smearing (Ross, Fabian \& Young 1999; Ballantyne, Ross \& Fabian 
2001; Done \& Gierli\'nksi 2006). 

\begin{figure*}
\begin{center}
\begin{tabular}{cc}
\leavevmode
\epsfxsize=0.45\textwidth \epsfbox{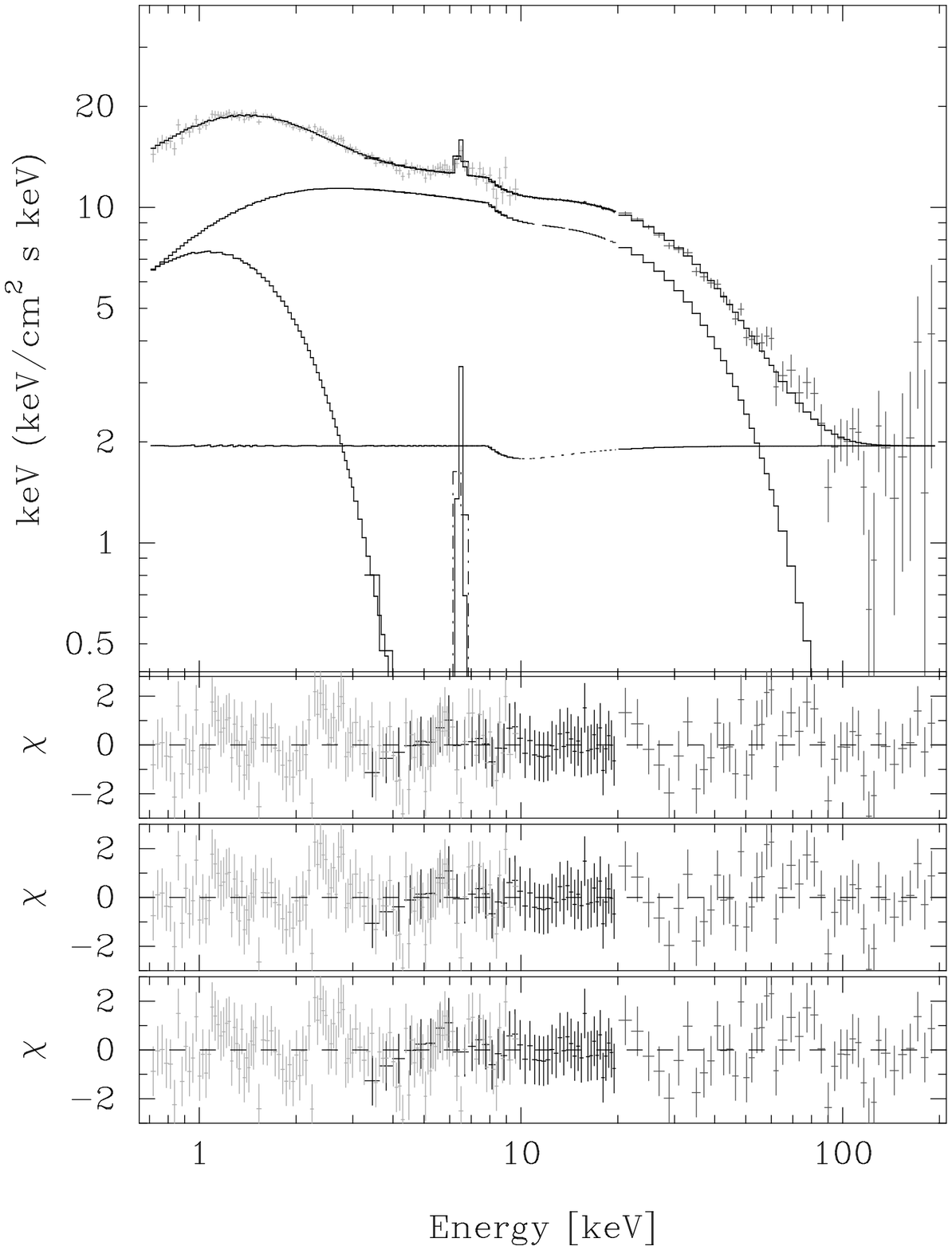} &
\epsfxsize=0.45\textwidth \epsfbox{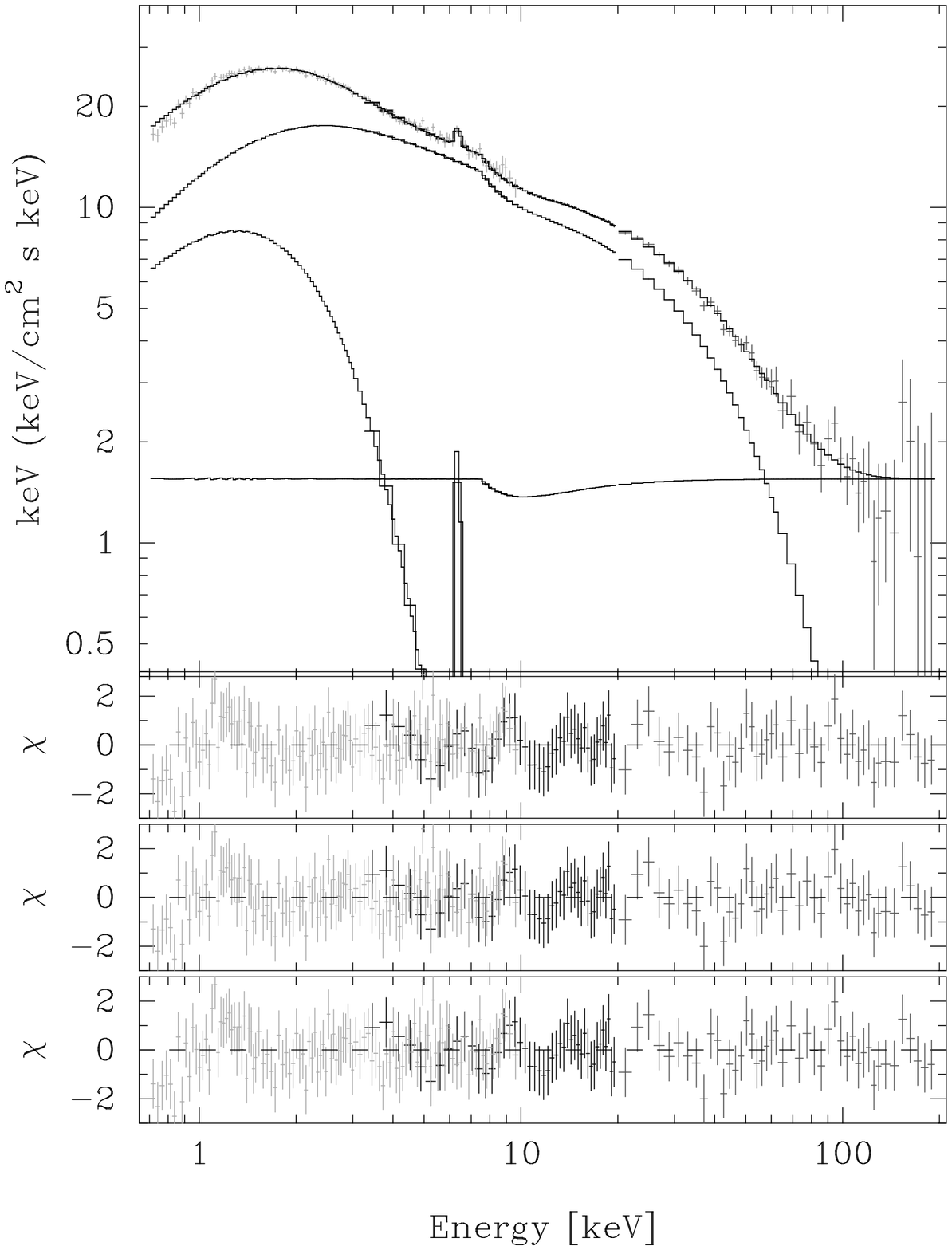}
\end{tabular}
\end{center}
\caption{Absorption corrected intrinsic spectra for each of the two
  observations described in the text. The left and right hand panels
  show the simultaneous ASCA-RXTE data taken before and after the
  peak, respectively. Both datasets are deconvolved using the best fit
  phenomenological model with SEG reflection, and the ASCA and HEXTE
  normalisations are set to match the PCA. The residuals are from this
  fit (SEG in Table A1), the {\sc dkbbth} fits (SEG in Table A2) and the
  {\sc dkbbfth} fits (SEG in Table A3), respectively, showing that all
  these different disc-corona models are a good description of the overall
  spectral shape.}
\label{fig:spec}
\end{figure*}

Fitting with better ionized reflection models which do include the
Compton scattering on the line and edge is not straightforward as
these models are tabulated for an illuminating power law continuum,
which is not appropriate for these data. Hence we attempt to quantify
the effect of these systematic model uncertainties in the reflected
emission by removing the 5-12~keV region (i.e. the
iron line and edge features which are sensitive to the disc
Comptonisation) from the spectral fits. We then refit the spectra with
ionization of the reflector fixed at $10^2$ (Xi2), $10^3$ (Xi3) and
$10^4$ (Xi4) in order to assess how much the derived disc luminosity
and temperature are dependent on this parameter, and so get an
estimate of the systematic model uncertainties (as in KD04).

We tabulate results from all free parameters in these fits in  
Table~\ref{tab:asca-rxte} in the Appendix. Uncertainties are given for 
$\Delta\chi^2=2.7$, and derived from a full error scan
where all other parameters are free to vary. We include the
derived disc luminosity, which is also subject to uncertainties in 
distance and inclination, but we stress that this is used only to 
compare with the disc luminosity derived using the {\em same} distance
and inclination. The reflected fraction is also dependent on the assumed
inclination, but this is not used further in the paper.  

Fig.~\ref{fig:spec}a and b shows each $\nu F_{\nu}$ spectrum,
deconvolved and corrected for absorption using the best fit SEG model
(phenomenological reflection), together with residuals. The two
spectra are qualitatively similar, with the disc emission being
strongly Comptonised. The only differences are that the uncomptonised
disc spectrum is slightly easier to identify in the first spectrum
than in the second, and the spectral index of the Comptonised emission
is slightly flatter. 
This immediately sets some constraints on the Comptonising geometry. The 
corona will scatter only a fraction $C_f [1-\exp(-\tau) ]$ where $C_f$ is the 
covering fraction of the corona over the disc and $\tau$ is the optical depth. 
Thus if either $C_f$ or $\tau$ were small then the disc spectrum would be 
above the Comptonised tail. Since it is not, the spectra in 
Fig.~\ref{fig:spec}a and b show that the majority of the disc emission is 
covered by an optically thick corona (see also the discussion in KD04). 

Both spectra also show two clear breaks within the
bandpass of the data, a high energy downturn at $\sim 30$~keV and a
low energy downturn at 1.2~keV and 1.8~keV for the first and second
datasets, respectively. We stress that the low energy downturn seen in
these data are {\em not} due to absorption, since the spectra in these
figures are absorption corrected, though it is dependent on the
assumption that the continuum seed photons are from the disc.  Our
derived column of $\sim 0.7\pm 0.2\times 10^{22}$ cm$^{-2}$ is
consistent within uncertainties with that of $0.8\pm 0.2\times
10^{22}$ cm$^{-2}$ derived from Chandra data (Miller et al 2003) and
is slightly less than the full column through our Galaxy in this
direction ($0.9\times 10^{22}$ cm$^{-2}$), as expected.

The existence of the low and high energy cutoffs within the observed
0.7--200~keV bandpass mean that these data cover the majority of the
accretion power output. Thus the total luminosity is well defined
(modulo uncertainties in the distance and inclination which cancel
when comparing to the disc luminosity derived from high/soft state
data), rather than being strongly affected by uncertainties associated
with extrapolating the spectral models outside of the observed range.

\subsection{Derived luminosity and temperature of the disc emission}

%%%% Figure 3 %%%%%%

\begin{figure*}
\begin{center}
\begin{tabular}{cc}
\leavevmode
\epsfxsize=0.45\textwidth \epsfbox{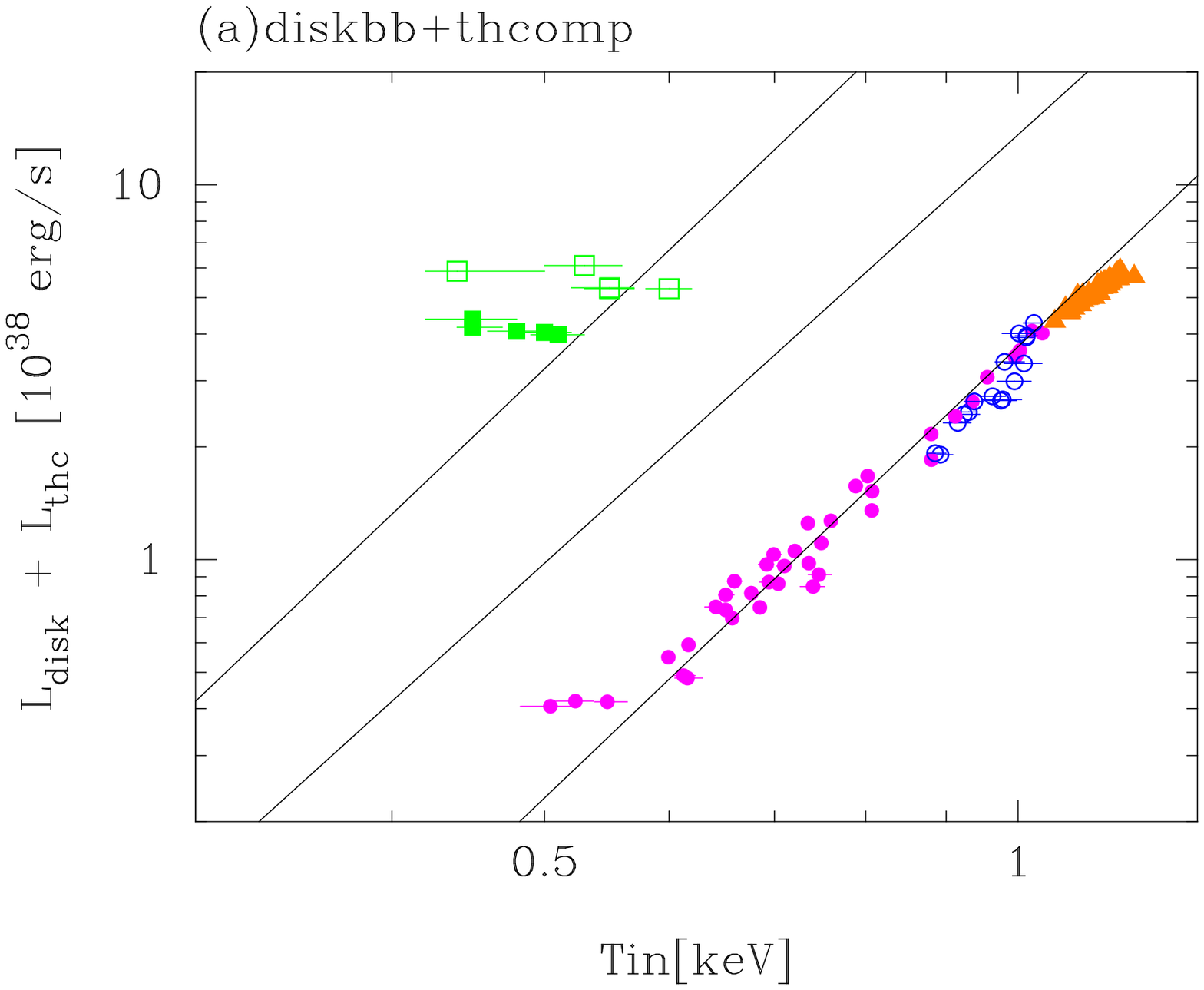}&
\epsfxsize=0.45\textwidth \epsfbox{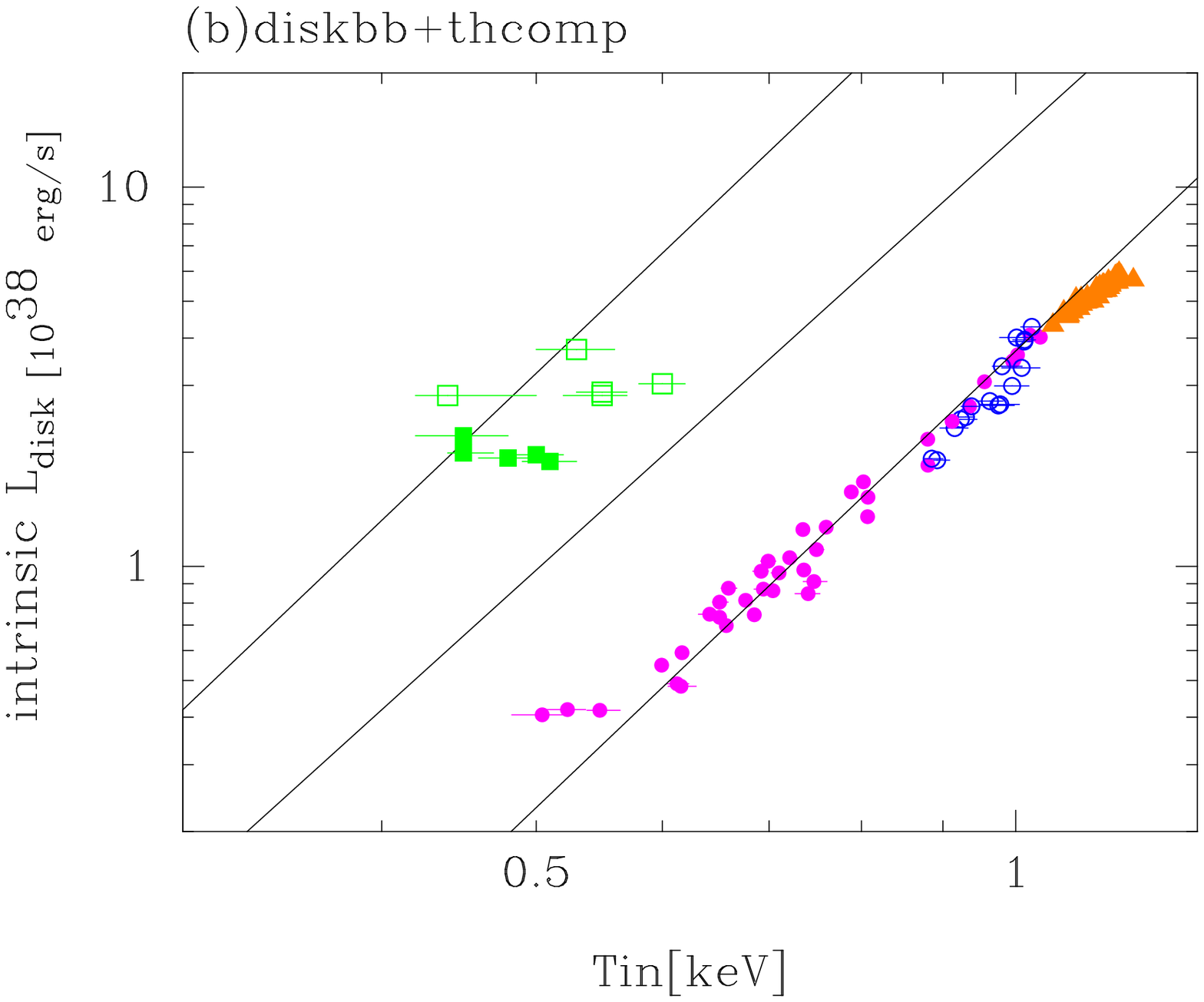}\\
\\
\\
\\
\epsfxsize=0.45\textwidth \epsfbox{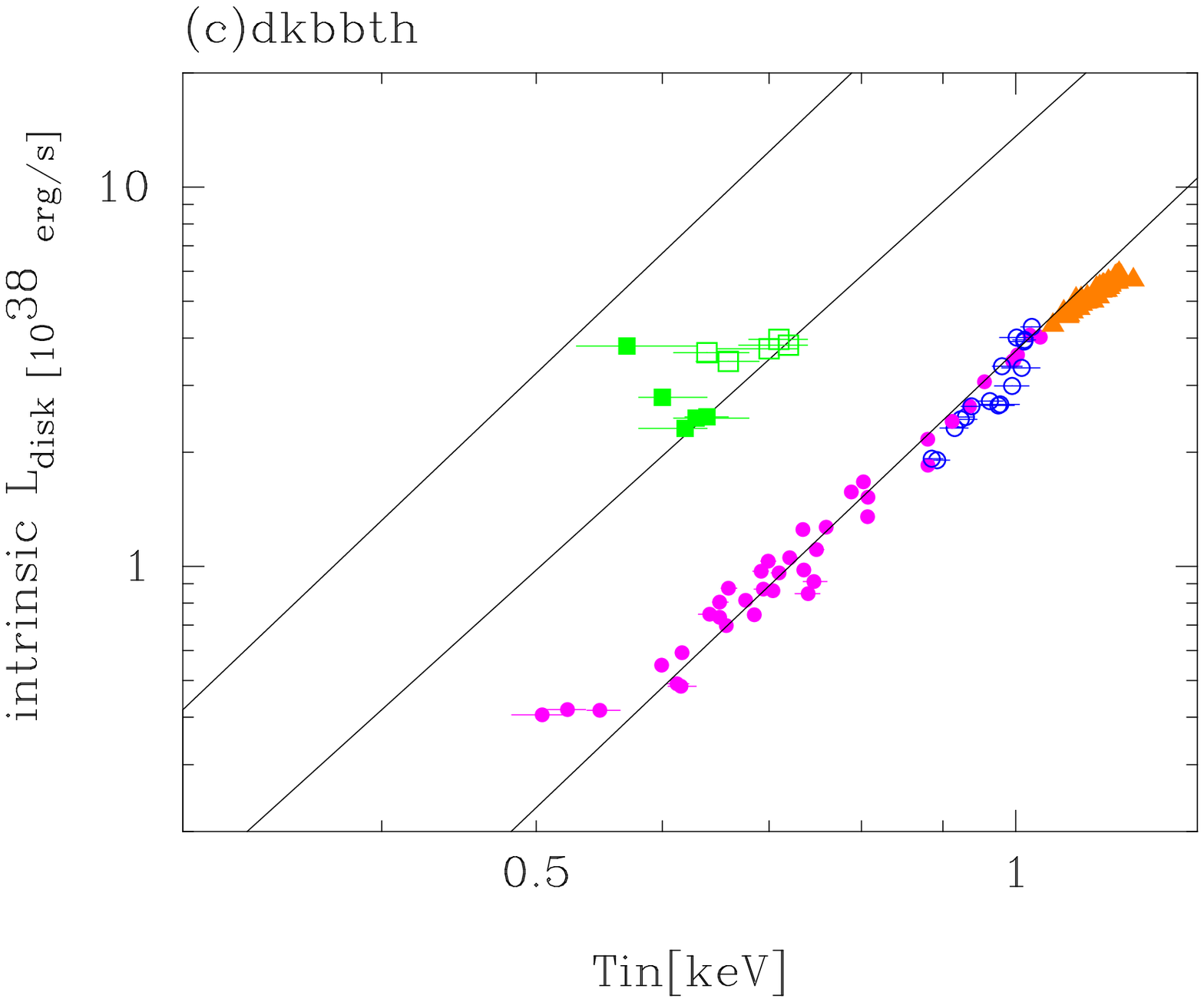}&
\epsfxsize=0.45\textwidth \epsfbox{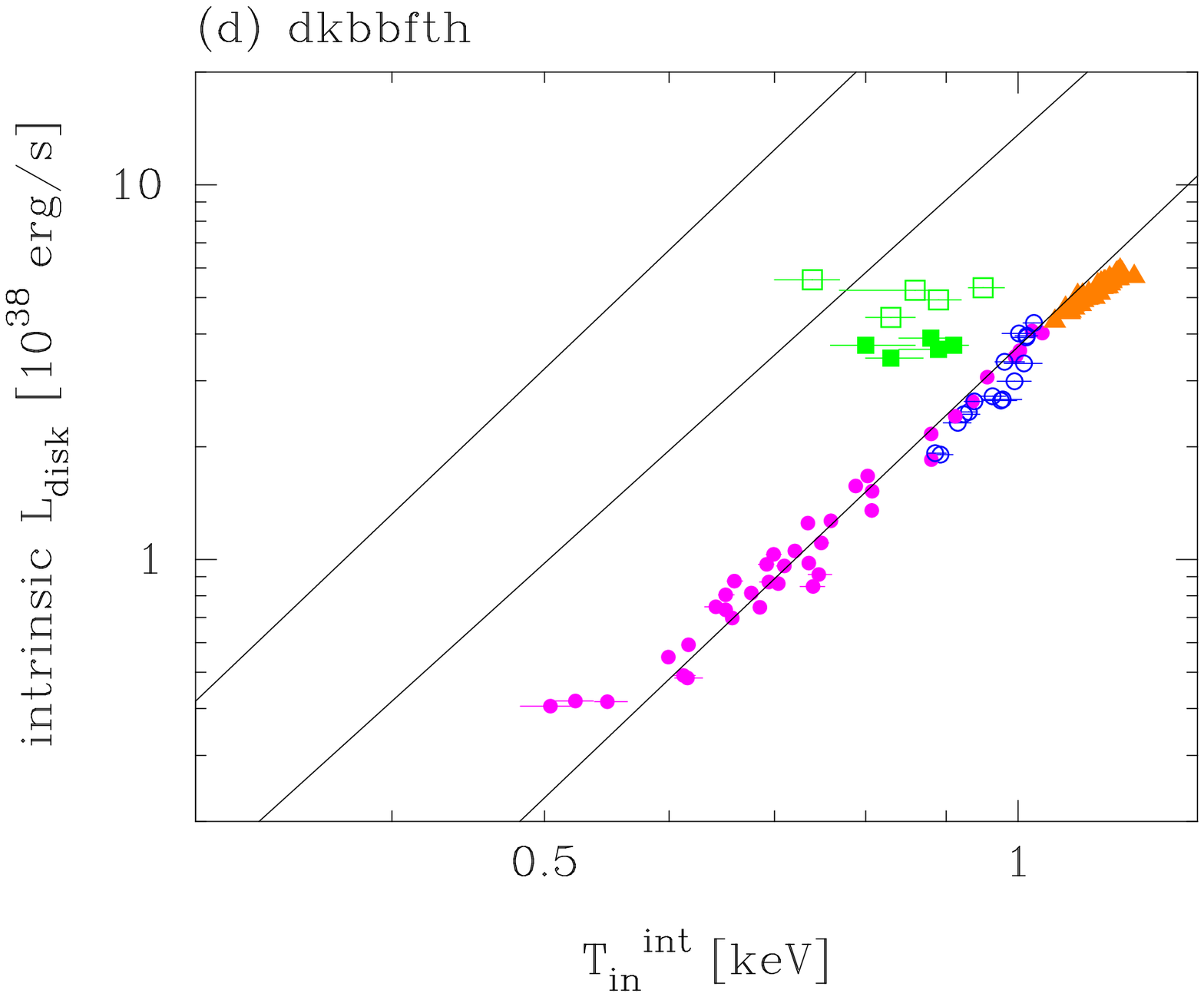}\\
\end{tabular}
\end{center}
\caption{Luminosity versus temperature plots for the disc. Filled
  circles and filled triangles show the data from the disc dominated
  states, while the open circles show the weakly comptonised very high
  state (KD04). Filled and open squares and open squares show the
  strongly comptonised very high state data from before and afer the
  peak, respectively. There are five points for each of these strongly
  comptonised spectra, corresponding to the five different reflection
  descriptions. The solid lines show $L\propto T^4$ for $r_{\rm
  in}=$59, 100 and 200~km. Panel a and b show the reconstructed disc
  emission from simple modelling of the spectra (Table A1),
  corresponding to a corona covering the whole disc (see
  Fig.~\ref{fig:sketch1}a). Panel a plots the observed disc
  temperature against the inferred disc luminosity estimated by
  $L_{disc}+L_{thcomp}$. This is a slight overestimate since
  Comptonisation boosts the seed photon luminosity by a factor $\sim
  1.3$ for these data. Panel b shows a more careful reconstruction of
  the disc luminosity using the photon numbers and inclination of the
  slab. Either way gives similar results, showing that the inferred
  disc luminosity and temperature is quite robustly estimated by these
  data, and that they imply the inner edge of the disc is much larger
  than that seen in the disc dominated spectra. Panel c shows the
  reconstructed inner disc temperature and luminosity for an inner
  disc-corona (Fig.~\ref{fig:sketch1}b). This is not very different to
  that found from the previous fits. By contrast, Panel d shows the
  results for an inner disc whose energetics are coupled to that of
  the corona (Fig.~\ref{fig:sketch1}c). The implied change in inner
  disc radius is now much smaller. }
\label{fig:lt}
\end{figure*}

Fig~\ref{fig:lt} shows the luminosity and temperature for the disc
dominated high/soft state observations of KD04 to compare with those
seen from the disc in the two VHS state spectra used here.  
Reconstructing the intrinsic disc emission is model dependent:
a Comptonising corona will distort the observed disc spectrum.
However, the Comptonised spectrum itself contains a
low energy rollover, directly imprinting the energy of the seed
photons on the boosted emission. Thus with broad
bandpass spectra, where the energy range of the expected disc
temperature is covered by the observations as in these ASCA-RXTE
data, the intrinsic seed photon temperature is recoverable. 

The seed photon luminosity is equally observable, despite the effects of 
Comptonisation shrouding the disc. Compton scattering increases the photon 
energy by on average $\Delta E/E \sim 4 \tau^2 (kT_{\rm e}/m_ec^2)$. Thus the 
Comptonised luminosity is related to the scattered disc luminosity simply by a 
factor $\sim (1+\Delta E/E)\sim 1.3$ for the steep Comptonised spectra observed 
here.  This scattered disc luminosity should simply be added to the observed 
disc luminosity to estimate the intrinsic disc luminosity of 
$(L_{disc}+L_{thcomp})/1.3$ Thus the intrinsic disc luminosity cannot be very 
much smaller than the total luminosity, which is robustly estimated by these 
data. This is shown plotted against observed disc temperature in 
Fig~\ref{fig:lt}a, and plainly shows a marked discrepancy with the constant 
radius/constant colour temperature correction observed from the disc dominated 
spectra.

A more careful correction for the effects of Compton scattering can be done by 
noting that this conserves photon number so the intrinsic disc luminosity can be 
simply estimated by increasing the {\em observed} disc luminosity by a factor 
$(aN_{comp}+N_{disc})/N_{disc}$ where $N_{comp}$ and $N_{disc}$ are the numbers 
of photons in the thermally Comptonised and disc spectral components, 
respectively, and $a$ are $2\cos i$ and $1$ for spherical and slab geometries 
(see Kubota \& Makishima 2004; KD04).  Panel b shows the reconstructed disc 
luminosity assuming a slab geometry and clearly shows that the inferred disc is 
very different to that seen in the high/soft state data.

Thus, while spectral fitting is almost always non-unique, the presence
of the low and high energy rollover in the observed bandpass mean that
we can make very robust estimates for the {\em intrinsic} disc
temperature and luminosity, giving a marked discrepancy with the
constant radius/constant colour temperature correction observed from
the disc dominated spectra. Disc dominated spectra at the same
luminosity have a temperature of $\sim1$~keV, so for a spectrum
dominated by Comptonisation we expect a low energy rollover from the
seed photon energy at $4kT_{\rm in}\sim4$~keV. This is in direct
conflict with the observation of the low energy rollover at $1.2$~keV
or $1.8$~keV which shows the seed photon temperature is $<0.5$~keV or
$<0.7$~keV. The inferred disc is highly luminous but with low
temperature, implying that the inner radius of the disc is a factor
$\sim 3$ higher in the VHS than in the disc dominated spectra.

Plainly, with these models the disc is {\em not} consistent with a
constant inner radius, constant emissivity, constant colour
temperature correction between the VHS and the disc dominated
state. At least one of these {\em must} be changing. An obvious change
is for the corona to affect the colour temperature
correction. Irradiation will change the vertical temperature structure
of the disc, but this should lead to an {\em increasing} temperature
for the upper layers, so an {\em increased} colour temperature
correction. Yet the data show a {\em lower} temperature for the
inferred luminosity. Similarly, conduction between the corona and disc
would also increase the disc temperature, as would an increase in the
stress at the inner disc boundary (Agol \& Krolik 2000). This robust
discrepancy between the disc structure in the disc dominated and VHS
motivates the more sophisticated modelling in the rest of the paper.

\section{Inner disc corona, separate energetics}

Here we build a simple model of an inner disc corona, and fit it to the data,
incorporating the effects 
of the geometry as well as the radiative constraints from Comptonisation. The 
corona is assumed to exist only at small radii (see Fig~\ref{fig:sketch1}b) so 
it removes only the hottest part of the disc spectrum by Comptonising it, while 
not affecting the spectrum of the cooler disc at larger radii. 
The aim is to test whether this more physically realistic model still requires 
the inferred increase in disc inner radius seen in the previous section.

\subsection{Models}

We assume that the corona is a slab of constant optical depth and
temperature on top of the disc, extending from $r_{\rm in}$ to $r_{\rm
c}$, while the disc itself extends from $r_{\rm in}$ to $1000r_{\rm
in}$ (see Fig.~\ref{fig:sketch1}b).  Each radius in the disc is
assumed to emit blackbody radiation, with the standard temperature
distribution of a multicolour {\sc diskbb} i.e. $T(r)=T_{\rm in}
(r/r_{\rm in})^{-3/4}$.  Where the disc is not covered by the Compton
slab then its emission is unaffected.  Otherwise, it is suppressed by
a factor $\exp(-\tau)$, where $\tau$ is determined from the
Comptonised spectrum using the observed $\Gamma$ and $kT_{\rm e}$,
assuming a slab geometry. The blackbody emission from the disc at this
radius is used as the seed photons for the Comptonising cloud, whose
spectrum is then normalised so that the total number of photons in the
Comptonised spectrum is equal to the number of photons scattered from
the blackbody disc emission at that radius i.e. $1-\exp(-\tau)$.

A more careful treatment would have to account for the different angle
dependences of the seed photons and comptonised emission - the number
of seed photons which emerge unscattered is more accurately
$\exp(-\tau/\cos i)$, but the number which are scattered into this
direction is {\em not} simply $1 - \exp(-\tau/\cos i)$ as photons
scattered into a given line of sight come from a range of initial seed
photon angles (see e.g. Pozdnyakov, Sobol \& Sunyaev 1983). There is
yet more complexity as about half the Compton scattered photons will
be directed down towards the disc. Those which are not reflected will
be reprocessed, adding to the intrinsic disc emission (Haardt \&
Maraschi 1993). We neglect all these effects in order to be able to 
approximately calculate the spectrum. 

%%%%%Figure 5%%%%%%%%%

\begin{figure*}
\begin{center}
\begin{tabular}{cc}
\leavevmode
\epsfxsize=0.45\textwidth \epsfbox{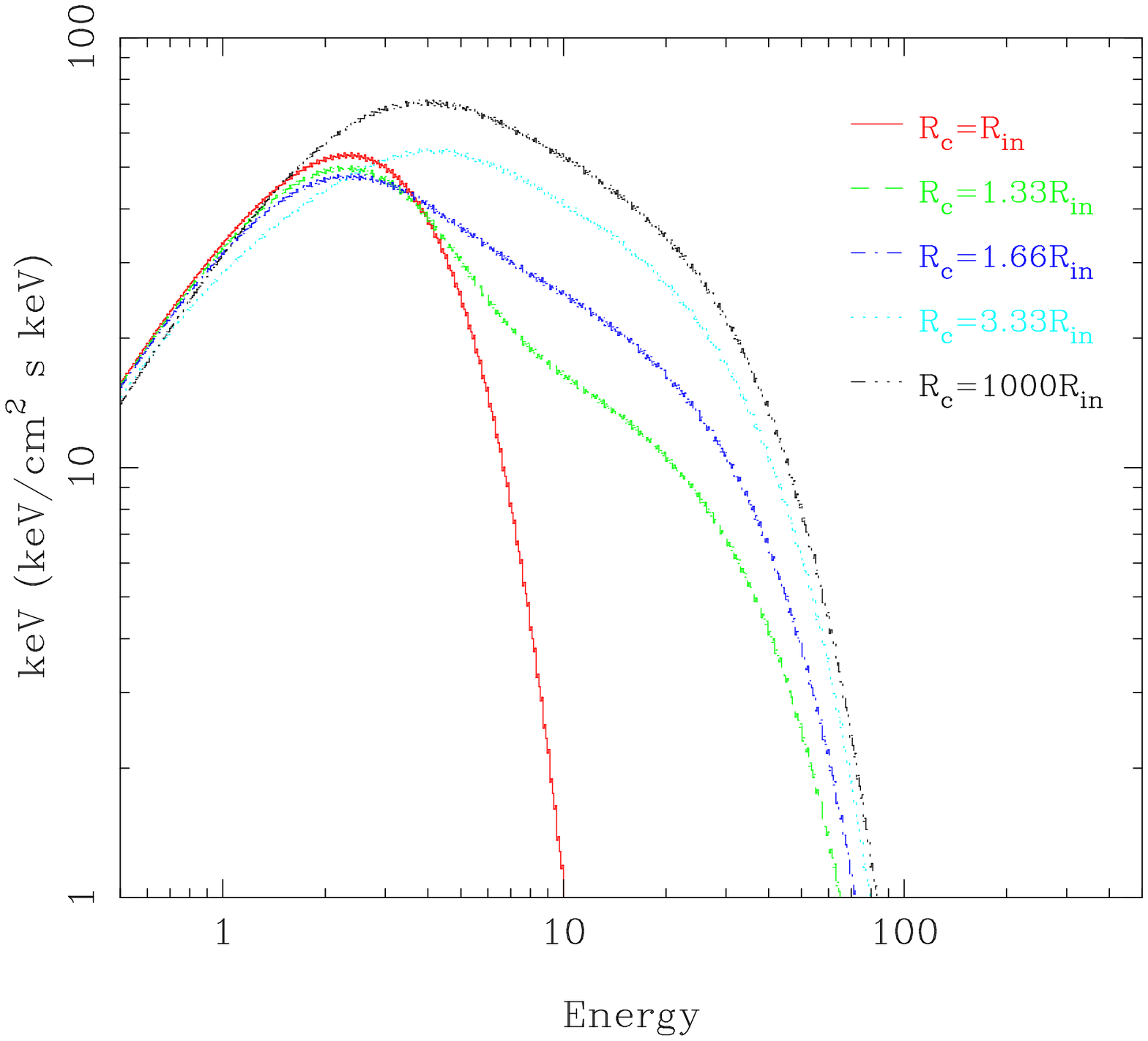} &
\epsfxsize=0.45\textwidth \epsfbox{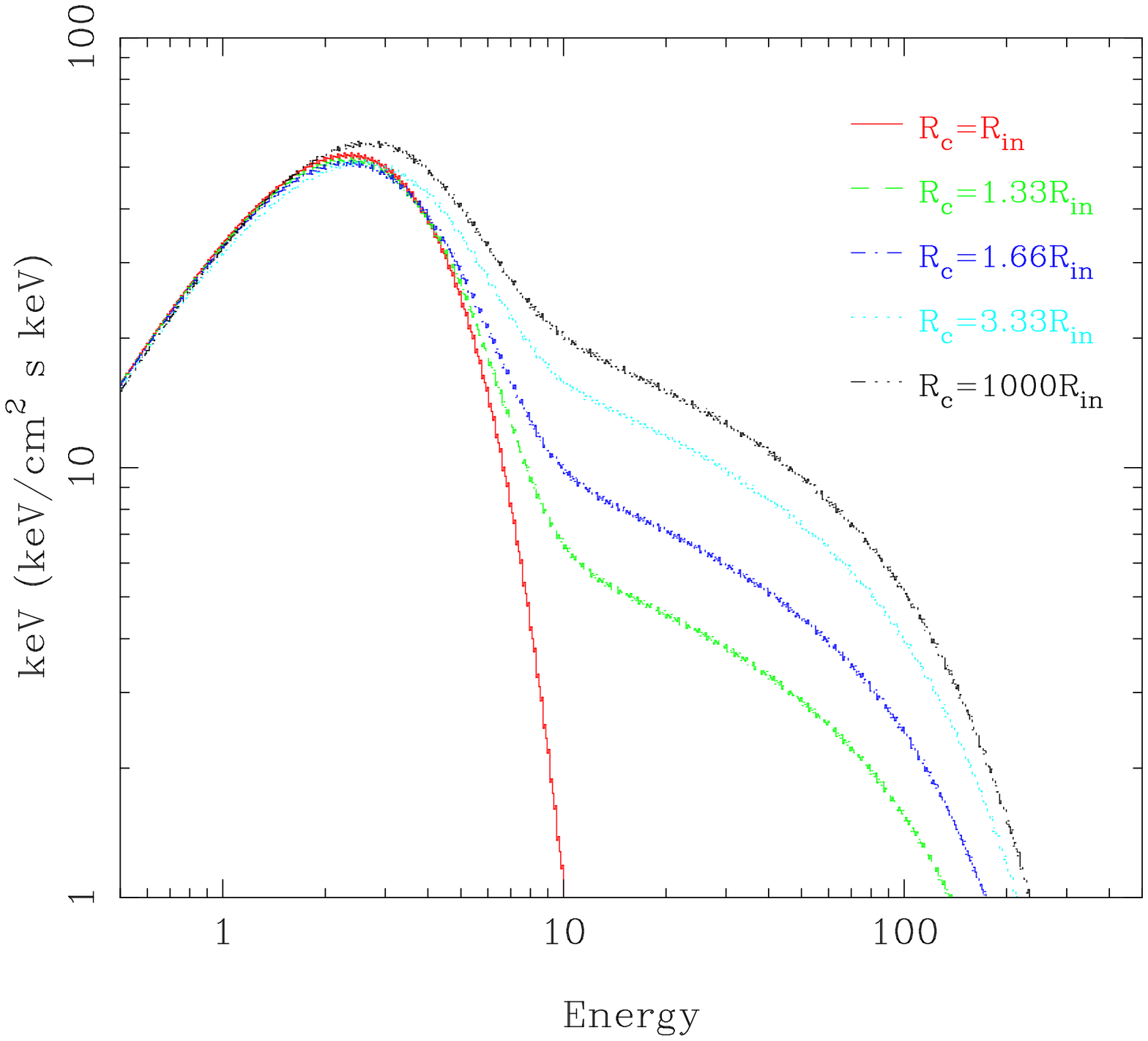}
\end{tabular}
\end{center}
\caption{An inner disc corona model for $r_{\rm c}=$ 1 (red, i.e. equivalent
to no corona), 1.33 (green), 1.66 (blue), 3.33 (cyan) and 1000 (black
i.e. equivalent to a corona which completely covers the disc) $\times
r_{\rm in}$ (see Fig.~\ref{fig:sketch1}b). 
The left panel shows the results for coronal
parameters close to those inferred from the data, with $\Gamma=2.3$ and
$kT_{\rm e}=10$~keV.  This corresponds to $\tau\sim 2$ for a slab corona. 
The right panel shows the effect of increasing the coronal
temperature to 100~keV.  To keep the same spectral index for higher
temperature means that the optical depth has to decrease to $\tau\sim
0.4$. Thus there is always a clear unscattered disc component (unlike
the data), and the intrinsic disc temperature of 1~keV is seen. }
\label{fig:model_dkbbth}
\end{figure*}

The shape of the spectrum is determined by the parameters of the
Compton corona ($\Gamma$ and $kT_{\rm e}$), the size of this corona
relative to the disc $r_{\rm c}/r_{\rm in}$, and intrinsic inner disc
temperature, $kT_{\rm in}$. The overall normalisation gives the apparent
disc radius, $r_{\rm in}$, for the assumed distance and inclination,
as in the {\sc diskbb} model. Even though the full {\sc diskbb}
spectrum {\em is not} seen, the model normalisation is that of the
{\em uncomptonised} disk. The model parameters for the disk component
are in effect what the disk would have looked like without the
Comptonising corona, and so can be directly compared with the disk
normalisation derived from the disk dominant high/soft state.

Fig.~\ref{fig:model_dkbbth}a shows the results for this model corona
parameters close to those implied by the VHS data i.e.  a corona with
$kT_{\rm e}=10$ keV and $\tau=2$ (which corresponds to a photon index
of the scattered flux of $\Gamma=2.3$ as the optical depth for a slab
geometry is half of that which gives the same spectrum for a
sphere). The inner disc temperature is set to 1~keV to match that seen
from the disc dominated spectra at similar luminosities (see
Fig.~\ref{fig:lt}).  The radius of the Comptonising region is
progressively increased over more and more of the disc, from $r_{\rm
c}/r_{\rm in}=1,1.33,1.66, 3.33$ and $1000$. This changes the spectrum
from a pure disc blackbody, to a more complex shape.  While this does
have the same number of free parameters as the previous model, it is
somewhat more constrained. The normalisation of the disc and
Comptonised components is not a completely free parameter, but is set
by $r_{\rm c}/r_{\rm in}$ and by the assumed slab geometry. In the
limit where the corona covers all the disc then the spectrum has the
same shape as that of a {\sc thcomp} model with disc blackbody seed
photons but with the addition of $\exp(-\tau)$ of the initial disc
blackbody spectrum which is unscattered by the corona.

The spectrum {\em always} shows the imprint of the seed photon energy
at $\sim 1$~keV. This is because of the physical requirements of the
model, firstly that the seed photons are from the inner disc with
standard emissivity, and secondly that the Comptonised spectrum has
the same number of photons in it as were removed from the disc
emission. A standard disc emissivity disc produces most of its
luminosity in the inner regions, and it is these inner regions which
are covered by the corona. The Compton energy boost is not large for
these coronal parameters ($\Gamma=2.3$, $kT_{\rm e}=10$~keV) so the
Comptonised photons, with their clear seed photon rollover, add to the
emission at similar energies to those of the seed
photons. Comptonisation conserves photon number, so the normalisation
of the spectrum is also similar. Thus the optically thick corona does
intercept most of the inner disc emission, but this Comptonisation
does not hide it as neither the energy nor number of photons are
changed substantially.

Figure~\ref{fig:model_dkbbth}b shows the same series of spectra
for increasing coronal radius, but with $kT_{\rm e}=100$~keV.
Though increasing the electron temperature increases the energy boost so
making it possible to move the photons away from their seed
energy, to get the same spectral index of $\Gamma=2.3$ from this
higher temperature plasma requires a lower optical depth. The corona
becomes optically thin so the spectrum now is clearly always two
component (disc plus Comptonisation), and the disc is always also
clearly at 1~keV.  Although only the inner disc photons are now
boosted far away from their seed photon energy, the low optical depth
means that not enough of them are Comptonised to hide the hottest disc
components even for a corona that extends over the whole disc. 

A comparison by eye of these models with the observed spectra in
Fig.~\ref{fig:spec} shows that the observed low energy rollover at
$\sim 1.8$~keV strongly constrains the seed photon energy even in an
inner disc corona model to be substantially less than 1 keV. Thus this
model is {\em not} going to easily change the conclusion of the
previous section that the
inner disc temperature is much lower than is seen at this luminosity
in the disc dominated high/soft states. These models of an inner disc
corona above a standard emissivity disc still require that the disc
inner radius in the VHS is substantially larger than in the high/soft
disc dominated states.

%%%% Fig.6 %%%%%%

\begin{figure*}
\begin{center}
\begin{tabular}{cc}
\leavevmode
\epsfxsize=0.45\textwidth \epsfbox{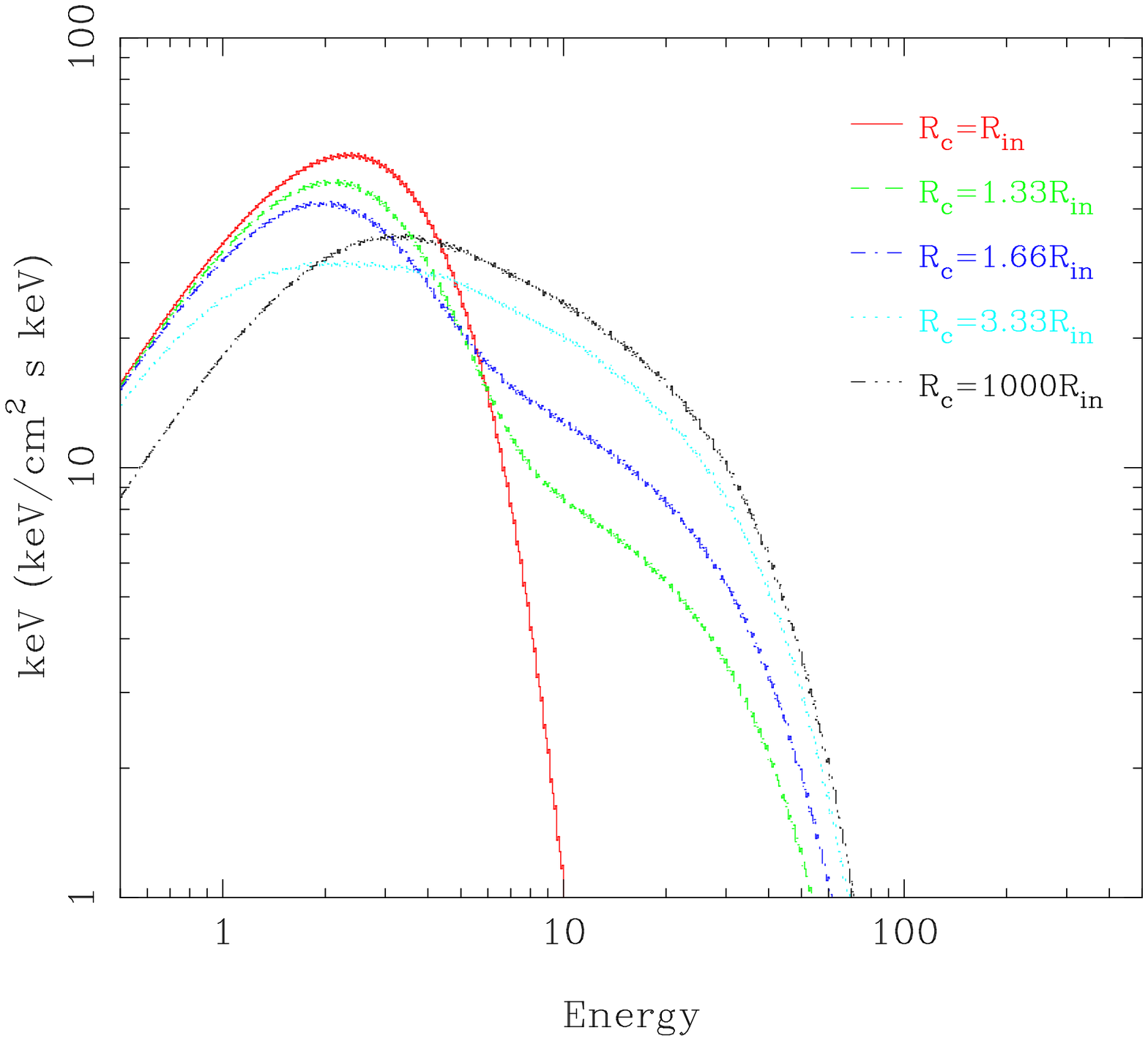} &
\epsfxsize=0.45\textwidth \epsfbox{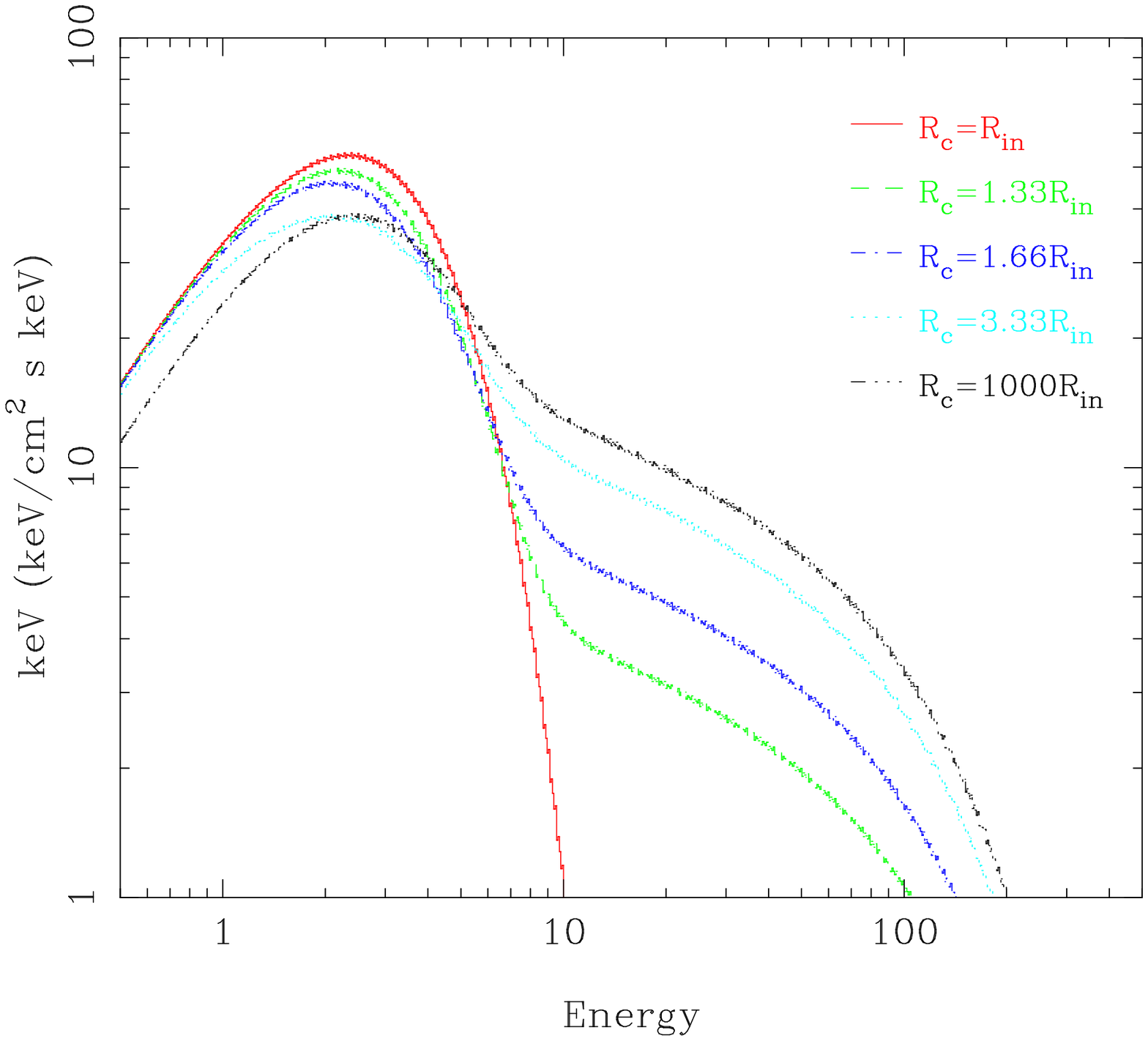}
\end{tabular}
\end{center}
\caption{As in Fig.\ref{fig:model_dkbbth}, but for an inner
disc-corona with coupled energetics (see Fig.~\ref{fig:sketch1}c). The
spectra left hand panel ($\Gamma=2.3$, $kT_{\rm e}=10$~keV) have
$f\sim 0.5$ for $r<r_{\rm c}$, while those in the right hand panel
($\Gamma=2.3$, $kT_{\rm e}=100$~keV) have $f\sim 0.3$.}
\label{fig:model_dkbbfth}
\end{figure*}

\subsection{Data fitting}

We quantify the argument above by replacing the separate {\sc diskbb}
and {\sc thcomp} components in the spectral fits with this new inner
disc corona model. We include the {\sc thcomp} reflection model
in the disc-corona code, to take into account 
illumination of the outer disc by the Comptonised spectrum. 
Again we use the same range of ionized reflection
descriptions as before in order to quantify the uncertainties which
arise from our modeling of this component.

First we fit this model to the simultaneous ASCA-RXTE spectra with the
coronal radius fixed at $r_{\rm c}/r_{\rm in}=1000$.  This model has
one less free parameter than the separate {\sc diskbb+thcomp} fits
shown in \S~4 because of the specific geometry (slab corona)
assumed. The observed spectral shape of the Comptonised emission
($\Gamma$ and $kT_{\rm e}$) set $\tau$ for the specific geometry, and
this fixes the ratio of normalisations of disc and Comptonised
emission. The fits are significantly worse, showing that our assumed
specific continuous slab corona geometry is not consistent with the
data. However, the inferred disc temperatures for the underlying
standard emissivity {\sc diskbb} spectrum are $\sim$~0.48~keV and
0.51~keV for the first and second data set, respectively. These are
much the same temperatures as for the previous fits (though with worse
$\chi^2$ due to the assumed geometry), illustrating the robustness of
reconstruction of the seed photon (intrinsic disc) temperature

We then repeat the fits with the coronal radius as a free parameter to
correspond to the geometry of Fig.\ref{fig:sketch1}b (again with all
five different reflection descriptions). While this formally gives the
same number of degrees of freedom as the separate {\sc diskbb+thcomp}
fits, the model is more constrained by the geometric and physical
assumptions. Nonetheless, the fits are indistinguishable in $\chi^2$
from those in \S~4, strongly supporting an inner disc-corona geometry
for these VHS spectra. Table A2 gives the results for the underlying
(i.e. intrinsic, before Comptonisation) {\sc diskbb} temperature and
luminosity. These are plotted in the middle panel of 
Fig.~\ref{fig:lt}b, together with the
points from the disc dominated spectra (KD04). Plainly, as shown
above, the ASCA-RXTE spectra still strongly require a very low
temperature disc which does not fall on the constant radius/constant
colour temperature correction line derived for the disc dominated
spectra. Again the inferred apparent inner disc radius 
is a factor 2 larger than that seen
in the high/soft disc dominated states.

\section{Inner disc corona, coupled energetics}

The previous model assumed that the disc emission is unaffected by the
presence of the corona. That is, the inner disc corona model assumes
that while the disc radiates efficiently at all radii, the coronal
mass flow only radiates within $r_{\rm c}$. However, this seems rather
artificial.  The corona dissipates some power, and this must derive
ultimately from the accretion flow.  The disc and coronal flows could
be kept separate in a continuous corona geometry
(e.g. Fig.~\ref{fig:sketch1}a) with some (approximately constant)
fraction of the power and/or mass accretion taking place via the
corona ($\dot{\rm Z}$ycki et al 1997, Haardt \& Maraschi 1993; Svensson \&
Zdziarski 1994).  The geometry of an inner disc-corona instead
implies that there is a {\em single} accretion flow at large radii,
$\dot{M}_{\rm disc, \infty}$ which then splits at $r_{\rm c}$ to form the
disc and corona (Fig..~\ref{fig:sketch1}c). Again, we build 
a model incorporating these concepts, and fit it to the data, to test
whether it still requires an increased inner disc radius.

\subsection{Models}

%%%%%%% Fig.7 %%%%%%%%%

\begin{figure*}
\begin{center}
\begin{tabular}{cc}
\leavevmode
\epsfxsize=0.45\textwidth \epsfbox{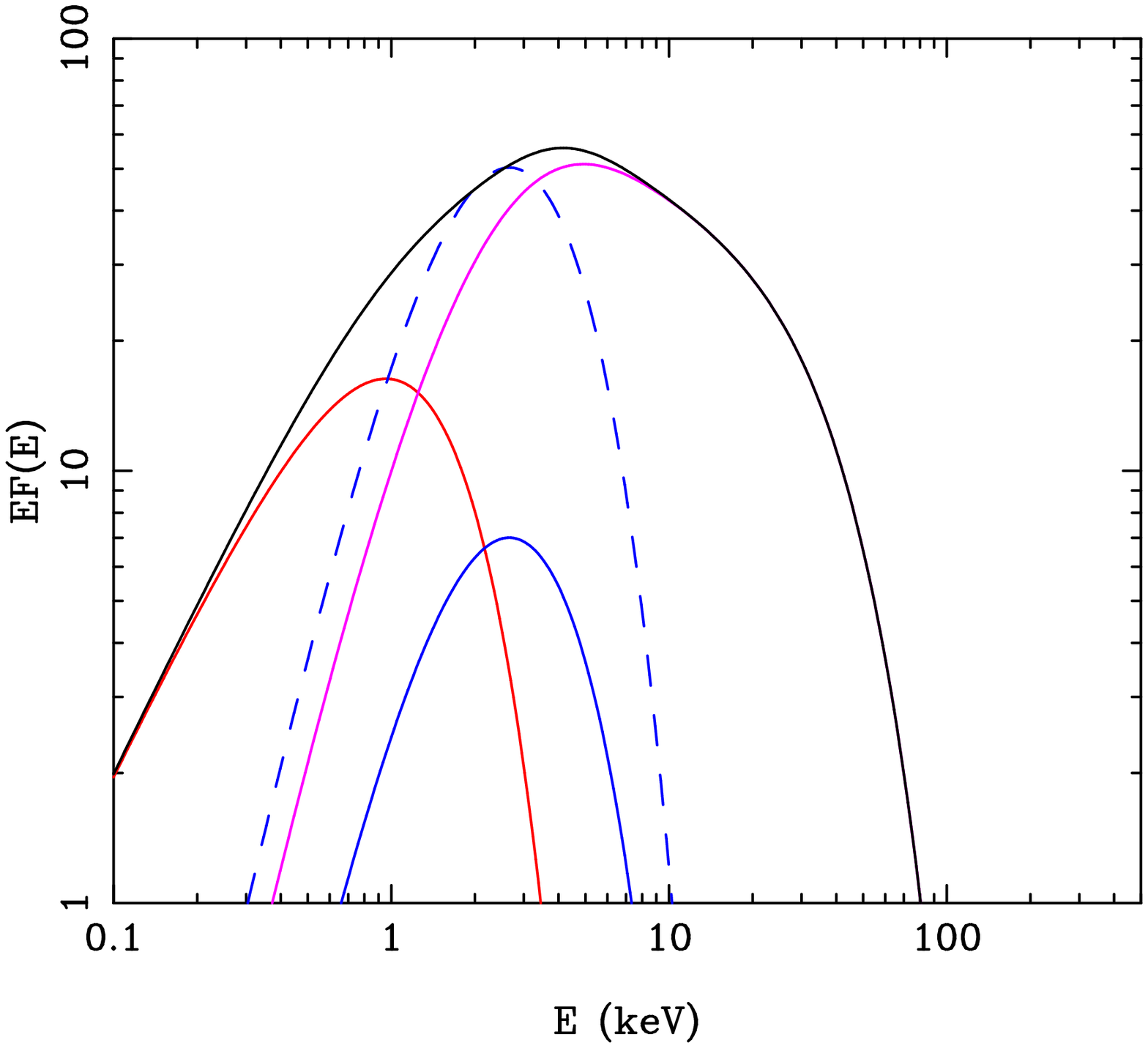} &
\epsfxsize=0.45\textwidth \epsfbox{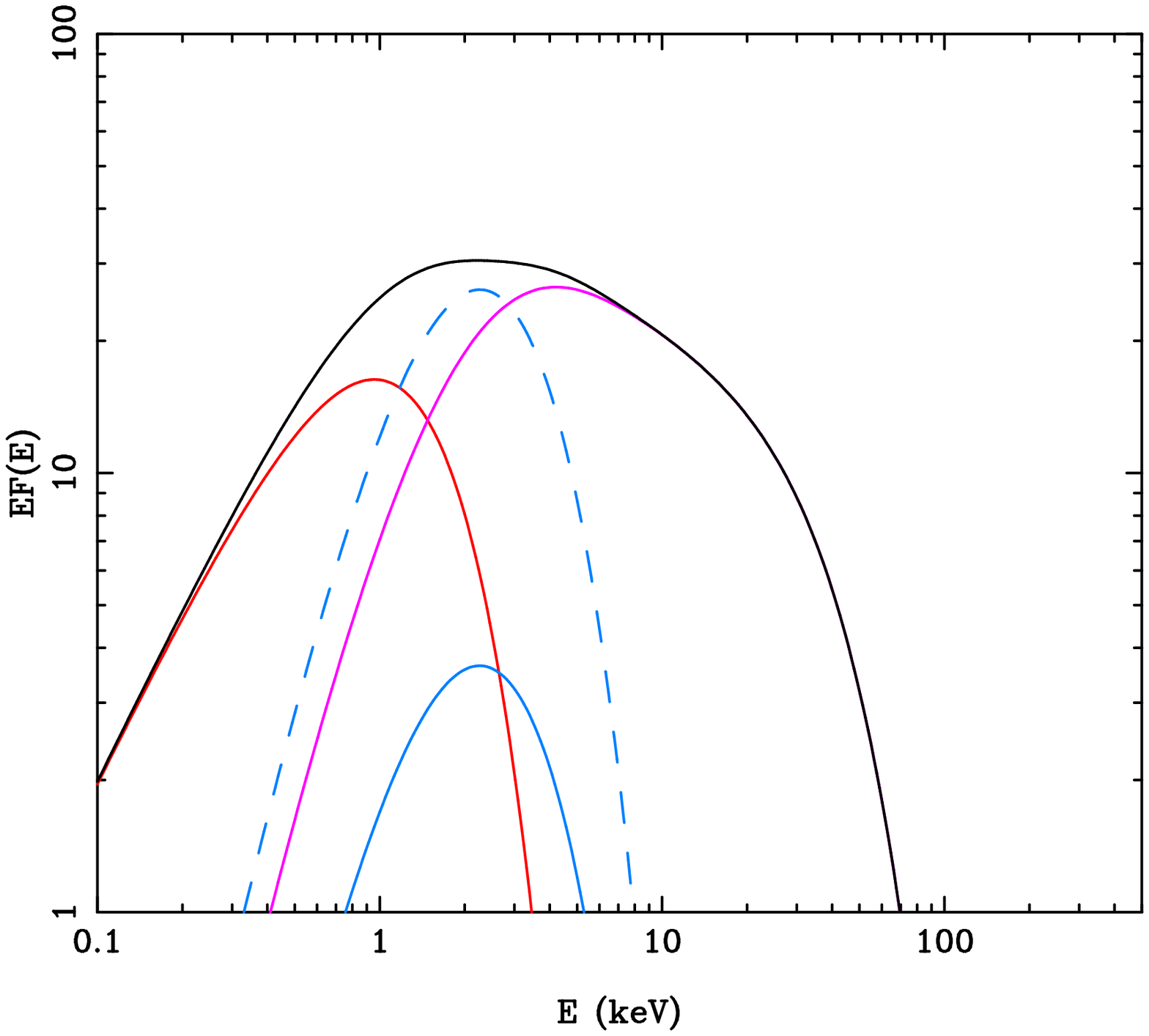}
\end{tabular}
\end{center}
\caption{The model spectra for $\Gamma=2.3$ and $kT_{\rm e}=10$~keV for $r_{\rm c}/r_{\rm in}=3.3$. The total spectrum
(black) is the sum of
the contribution from the outer disc (red), 
uncomptonised inner disc (blue line), and
comptonised emission (magenta). The dashed blue line shows the
intrinsic (before Comptonisation) inner disc emission. 
The left panel shows the uncoupled
disc corona, while the right panel shows the coupled
system.  There is a marked decrease in luminosity in the coupled
system due to the assumed emissivity
(compare the right hand panels of
Fig.~\ref{fig:sketch1}b and c), but more importantly 
there is also a marked decrease in the low energy rollover 
seen in this spectrum. This is 
mostly due to the decrease in emissivity of the 
inner disc emission. This lowers the normalisation of the comptonised
spectrum, so increasing the 
relative contribution of the low temperature, outer disc. 
}
\label{fig:dkbbth_fth_nth}
\end{figure*}

We use the specfic disc-corona coupling model of 
Svensson \& Zdziarski (1994), which assumes most of the mass accretion rate is
still carried by the disc, but where a fraction $f$ of the accretion
power is dissipated in the corona. The power dissipated in the
optically thick disc is discontinuous at $r_{\rm c}$, where it drops
from $\dot{M}_{disc,\infty}$ by a factor $(1-f)$. This implies the
disc temperature is also discontinuous, dropping to $(1-f)^{1/4}$ of
that expected from $\dot{M}_{disc,\infty}$ (see Svensson \& Zdziarski (1994) calculate the effect on
the disc in the limit in which most of the mass accretion rate is
still carried by the disc, but where a fraction $f$ of the accretion
power is dissipated in the corona. The power dissipated in the
optically thick disc is discontinuous at $r_{\rm c}$, where it drops
from $\dot{M}_{disc,\infty}$ by a factor $(1-f)$. This implies the
disc temperature is also discontinuous, dropping to $(1-f)^{1/4}$ of
that expected from $\dot{M}_{disc,\infty}$. This is shown
schematically in Fig.~\ref{fig:sketch1}c.

We incorporate these ideas into our inner disc-corona code to predict
the spectrum from a coupled disc-corona system.  The code starts with
the standard emissivity disk to produce an initial guess at the seed
photon luminosity and temperature and calculates the Comptonised flux
as before (using $\Gamma$ and $kT_{\rm e}$ to determine $\tau$, and
scattering a fraction $1-\exp(-\tau)$ of the disc photons into the
Comptonised emission). The total Comptonised flux from all radii is
used to calculate a (radially averaged) value of $f$ which is used to
re-calculate the disk emissivity and temperature, acting as seed
photons for the Comptonising corona, giving a slightly different value
of $f$. The code iterates once more around this loop in order to reach
a self-consistent solution. The parameters of the model are the same
as for the uncoupled inner disc-corona system described above. Again,
the inner disc temperature is that of the {\em uncomptonised} disk
i.e. the {\sc diskbb} which would have been seen {\em without} the
corona present. For the uncoupled disk-corona this is simply the
underlying disc emission, whereas for the coupled disk-corona the
underlying disc emission is distorted from that of a {\sc diskbb} by
the effect of the corona draining energy from the disk. The
'temperature' which parameterizes this coupled disk-corona systems is
the innermost disc temperature, $T_{\rm in}^{\rm int}$, which would be
seen in the limit of $f=0$, {\em not} the inner disc temperature used
in making the model which is $(1-f)^{1/4}T_{\rm in}^{\rm int}$.

Fig.~\ref{fig:model_dkbbfth}a and b shows the spectra obtained from
such models with  the same intrinsic disc ($T_{\rm in}^{\rm int}=1$~keV)
and coronal parameters as for the separate
disc-corona shown in Fig.~\ref{fig:model_dkbbth}a (where $\Gamma=2.3$,
$kT_e=10$~keV, implying $f\sim 0.5$ in the corona) and b (where
$\Gamma=2.3$, $kT_e=100$~keV, implying $f\sim 0.3$ in the corona).

Fig.~\ref{fig:dkbbth_fth_nth} illustrates the differences between the
coupled and uncoupled coronae by showing the individual spectral
components for a corona with $\Gamma=2.3$, $kT_e=10$~keV and $r_{\rm
c}/r_{\rm in}=3.3$.  Both spectra have the same $\dot{M}_{{\rm disk},
\infty}$, and assume the same inner disc radius (so without the corona
both would give the same disc temperature of $T_{\rm in}=T_{\rm
in}^{\rm int}= 1$~keV) but there is much less total power in the
coupled disc-corona system. This is as expected, as there is the
additional coronal emissivity in the uncoupled disc-corona model.
More importantly, the low energy rollover in the spectrum which is
used to indicate the seed photon temperature is much lower in the
coupled disc-corona. The reason for this is {\em not} primarily due to
the reduction in temperature of the inner accretion flow under the
corona. This is only a factor $(1-f)^{1/4}$ lower than $T_{\rm
in}^{\rm int}$ i.e. 0.85~keV, so the low energy rollover in the
Comptonised spectra is {\em not} significantly smaller ($4kT_{\rm
in}$=3.4~keV as opposed to 4~keV; KD04, magenta line). The major
reason for the lower energy rollover in the total temperature is the
reduction in {\em emissivity} of the inner disc in the coupled corona
system.  The dotted blue lines show the intrinsic inner disc emission,
while the solid blue lines show the emergent, uncomptonised disc
emission. The intrinsic disc emission is a factor $(1-f)$ weaker in
the coupled disc-corona, so the Comptonised spectrum is also a factor
$(1-f)$ weaker. A standard emissivity {\sc diskbb} has maximum
luminosity from the inner disc components, hence the Comptonised
emission (with its high seed photon rollover temperature) is generally
{\em dominant}. For the coupled disc-corona, the reduction in
emissivity under the corona means that the luminosity of the corona
can be {\em less} than the luminosity of the outer disc. The
Comptonised emission in the coupled system still contains a fairly
high seed photon rollover, at $(1-f)^{1/4}T_{\rm in}^{\rm int}$, but
this can be masked by the much lower temperature rollover from the
{\em outer} disc emission at $\sim T_{\rm in}(r_{\rm c}/r_{\rm
in})^{-3/4}\sim 0.4$~keV.  The major effect of the coupled energetics
is to decrease the fraction of energy emitted in the inner
disc-corona, so enhancing the relative importance of the cooler, outer
disc emission.

\subsection{Data Fitting}

We refit the data with this coupled disc-corona model, in which the
disc emissivity {\em changes} from that of a standard disc. Again we
use all 5 different descriptions of reflection to quantify the effect
of systematic uncertainties in spectral modelling. Details of the fits
to the data are given in Table A3, and again the models give equally as
good a description of the data as either the uncoupled inner disc/slab
corona (\S~5), or the continuous disc/corona system (\S~4). 

The most obvious change from the coupled energetics is that the
inferred inner disc radius is much smaller than in the previous fits.
This is due to the same combination of two effects as given above
(\S~6.1) i.e. firstly and most importantly the increased relative
strength of the low temperature emission from the outer
(uncomptonised) disc, and secondly the slight reduction in temperature
of the inner disc due to the energy drained from it to power the
corona.  The right hand panel of Fig.~\ref{fig:lt} shows the the
derived luminosity and temperature of the disc {\em without} its
corona, $T_{\rm in}^{\rm int}$, compared to the high/soft state
luminosity-temperature data. The VHS points now lie much closer to
those from the disc dominated states than before (left and middle
panels of Fig.~\ref{fig:lt}) though formally the statistical
uncertainties show that the 20~\% increase in disc radius in the VHS
is significant (assuming a constant colour temperature correction).
However, this is probably completely compatible with a constant disc
radius given the systematic modelling uncertainties e.g. our
assumptions about the shape of the corona (slab, with constant $\tau$
and $T_e$), the form of disc-corona coupling (all $\dot{M}$ through
the disc, as opposed to some fraction of the accretion also taking
place via the corona) and neglect of self-consistent reprocessing of
the illuminating coronal flux by the disc.

\section{Summary of spectral fits}

The VHS spectra seen in some high luminosity Galactic black holes are plainly 
strongly affected by Compton scattering. The most extreme VHS spectra no longer 
have a clear disc component, separate from the Compton scattered flux, but 
instead show a smooth continuum shape. The clear observational inference is that 
the radii which produce the majority of the disc luminosity must be covered by 
optically thick material.  

Assuming the inner disc is covered by an optically thick corona obviously makes 
it difficult to observe the properties of the inner disc, though the intrinsic 
disc luminosity and temperature can still be reconstructed from 
broad bandpass spectra. The temperature can be determined from the low energy 
rollover in the Compton scattered spectrum, while the disc luminosity is related 
to the Compton scattered luminosity simply through the energy boost from 
Comptonisation. 

We use the two
simultaneous ASCA-RXTE (0.7--200~keV) observations of the extremely
Comptonised VHS spectra from XTE~J$1550-564$ to investigate the
intrinsic disc emission.  As shown by KD04, this emission is {\em not}
consistent with that seen in the disc dominated state from the same
object. The temperature is much lower than that seen during the disc
dominated states of the same luminosity. 

We build a simple model of a slab corona over a disc to show that this 
conclusion holds irrespective of whether the corona covers the whole disc, or 
only the inner portion. The data strongly require that either the disc radius 
has increased, or the disc emissivity has decreased (or both).
We sketch each of these possible VHS geometries in
Fig.~\ref{fig:sketch2}a-c, respectively.  If the disc does truncate,
the accretion flow should still extend down to the last stable orbit
unless there is a powerful outflow (wind/jet) at this point.  Thus we
show a continuous corona inward of the truncated disc as being the
likely geometry in Figs.\ref{fig:sketch2}a and c.

The one common denominator for {\em all} the geometries in
Figs.~\ref{fig:sketch2} is that the disc is {\em strongly} affected by
the Comptonising corona. The coronal flow is {\em not} some separate
phase of accreting material, e.g. a low angular momentum flow. Instead
it {\em must} couple in some way to the disc.

\section{Additional constraints from the data}

\begin{figure}
\begin{center}
\leavevmode
\epsfxsize=0.45\textwidth \epsfbox{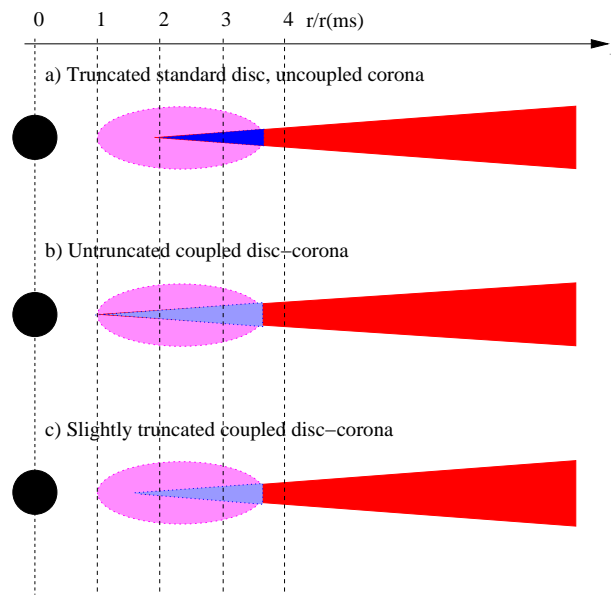}
\end{center}

\caption{Possible geometries for the VHS, with radii for the disc and
  corona in a) and c) taken from the average of the Xi1, Xi2 and Xi3
  model fits to the spectrum after the peak with {\sc dkbbth} (Table
  A2) and {\sc dkbbfth} (Table A3), respectively.  }
\label{fig:sketch2}
\end{figure}

There is additional information in the spectral and
variability properties of these data which can give further
constraints on the geometry. Firstly, the overall shape of the
Comptonised spectrum is steep ($\Gamma>2$), so the average energy gain is rather
small. This means that the seed photon luminosity is similar to the
luminosity emitted by the electrons i.e. to the heating rate of the
electrons. This is easy to arrange in the corona over the disc since
the disc provides copious seed photons. However, it is much more
difficult in the geometry of Fig.~\ref{fig:sketch2}a, with the
suggested continuation of the corona over a large range (factor 2) in
radius down to the last stable orbit. The majority of the
gravitational potential energy is dissipated in a region where there
is no disc underneath, so the only seed photons for Compton cooling
are a small fraction of those from the disc at larger radii. The
combination of less Compton cooling and more electron heating would
surely drive the coronal temperature to much higher values, leading to
a much harder spectrum (as in the low/hard state).  The smaller disc
truncation implied by the geometry of Fig.~\ref{fig:sketch2}c is much
less of a problem, as the inner coronal flow is much smaller, so is
energetically much less important.

The QPO behavior can also give some clues to the geometry, though these are 
somewhat ambiguous as their origin is not yet well understood. The strongest QPO 
is the one at low frequency (LF QPO: type C) and this increases in frequency as 
the spectra make a transition from the low/hard state to one of the soft states 
(e.g. van der Klis 2000). This indicates that the characteristic radius picked 
out by the variability is decreasing with decreasing strength of Comptonisation 
in the spectrum (e.g. di Matteo \& Psaltis 1999). Qualitatively this can be 
explained if the disc is truncated in the low/hard state, and that its inner 
edge moves inwards towards the last stable orbit during the transition. The fact 
that the LF QPO is not at the highest observed frequency in these spectra then 
indicates that the disc is still slightly truncated in these data, favouring the 
geometry of Fig.~\ref{fig:sketch2}c over that of Fig.~\ref{fig:sketch2}b where 
the disc radius is already fixed at the last stable orbit. 

We quantify the extent of disc truncation using the specific model of Stella, 
Vietri \& Morsink (1999), in which the LF QPO frequency, $\nu_{QPO}\propto r^{-
0.33}$, is set by Lense--Thirring precession of vertical perturbations of the 
innermost edge of the disc. The LF QPO's (Type-C) are at 2.38~Hz and 3.90~Hz for 
the 1st and 2nd observations respectively, while in the weak VHS data, these are 
found at 5.9-- 6.5~Hz as Type-B QPOs (Remillard et al.~2002).  The ratio of QPO 
frequencies between those seen in these strong VHS, and the most weakly 
Comptonised VHS is 2.6 and 1.6, implying a change in radius in the relativistic 
precession of about a factor 1.4 and 1.2 for the first and second dataset, 
respectively. This assumes that the weakly comptonised VHS has disc close to the 
last stable orbit (KD04). These estimates from the QPO's are startlingly close 
to the radii derived from our coupled disc- coronal fits, which imply a disc 
inner radius of $\sim 1.35- 1.1\times$ and $1.3- 1.1\times$ that inferred from 
the high/soft state data. However, the major problem with such models is that 
the LF QPO is seen in the spectrum of the corona, not of the disc (van der Klis 
2000; Gilfanov, Revnivtsev \& Molkov 2003; Sobolewska \& $\dot{\rm Z}$ycki 
2006). The corona may be driven at this frequency by the disc, or the LF QPO 
might represent a mode of the corona itself, in which case there are no 
constraints on the geometry. 

The high frequency (HF) QPO seems initially more constraining as the rapid 
timescale implies that it originates close to the black hole. This almost 
certainly arises from a resonance, as it appears as two separate features in a 
3:2 ratio (see e.g. the review by Remillard 2005). If this is the parametric 
resonance between vertical and radial epicyclic modes of a disc in strong 
gravity then this requires that the HF QPO comes from the innermost stable orbit 
of the disc around a high spin black hole (Abramowicz \& Kluzniak 2001). This 
model would also require that the inner disc can be seen directly, in apparent 
conflict with the basic model presented here of an optically thick inner disc 
corona. However, we note that the HF QPO's are weak, and are not seen in our 
datasets (Remillard et al 2002). Instead they are detected in VHS spectra from 
this source where the tail is not as strong as seen here (Remillard et al 2002; 
KD04; Remillard 2005). It is possible that the HF QPO comes from the small fraction 
$\exp(-\tau)$ of emission from the disc which is able to escape directly 
(uncomptonised), so is absent in our data simply because these lines of sight 
are rarer when the corona is more optically thick and/or covers more of the 
inner disc. However, it is also possible that the HF QPO represents a resonance 
of the corona, in which case there are no visibility constraints (or constraints 
on black hole spin in the specific model of Blaes, Arras \& Fragile 2006).

The only observation which seems in direct conflict with all models of an 
optically thick inner disc corona is that of the strong, extremely broad iron 
line claimed for the second set of ASCA data used here (Miller et al 2004). 
Again, a small component, $\exp(-\tau)$ of the highly skewed line from the inner 
disc could be seen directly, but this is much weaker than the feature observed. 
Either the X-ray tail in the VHS is not formed by Comptonisation, or the 
spectral residuals around the iron line region do not indicate a highly 
relativistic line from the inner disc. 

\section{Alternatives to Comptonisation}

\begin{figure*}
\begin{center}
\begin{tabular}{cc}
\leavevmode
\epsfxsize=0.45\textwidth \epsfbox{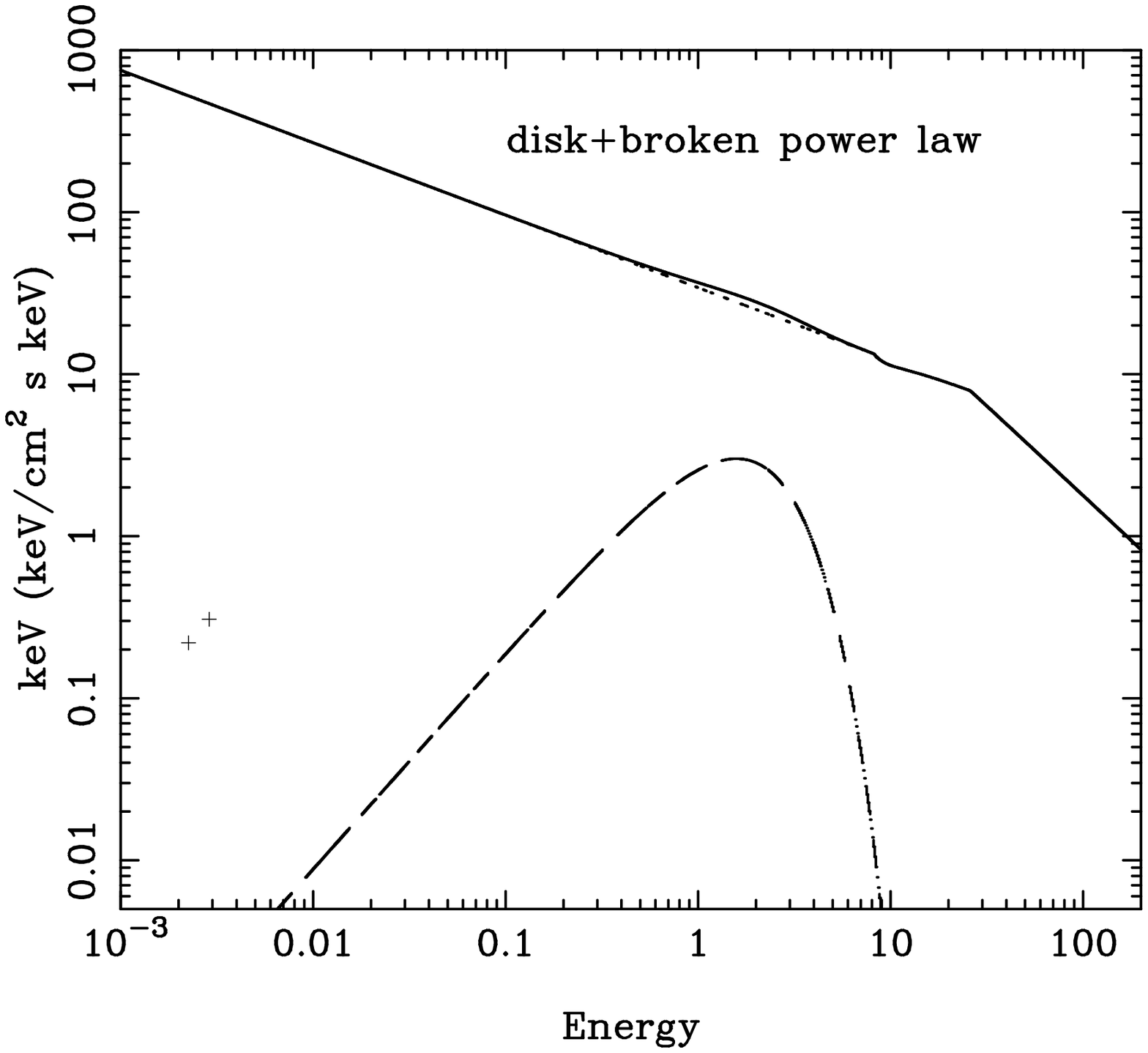} &
\epsfxsize=0.45\textwidth \epsfbox{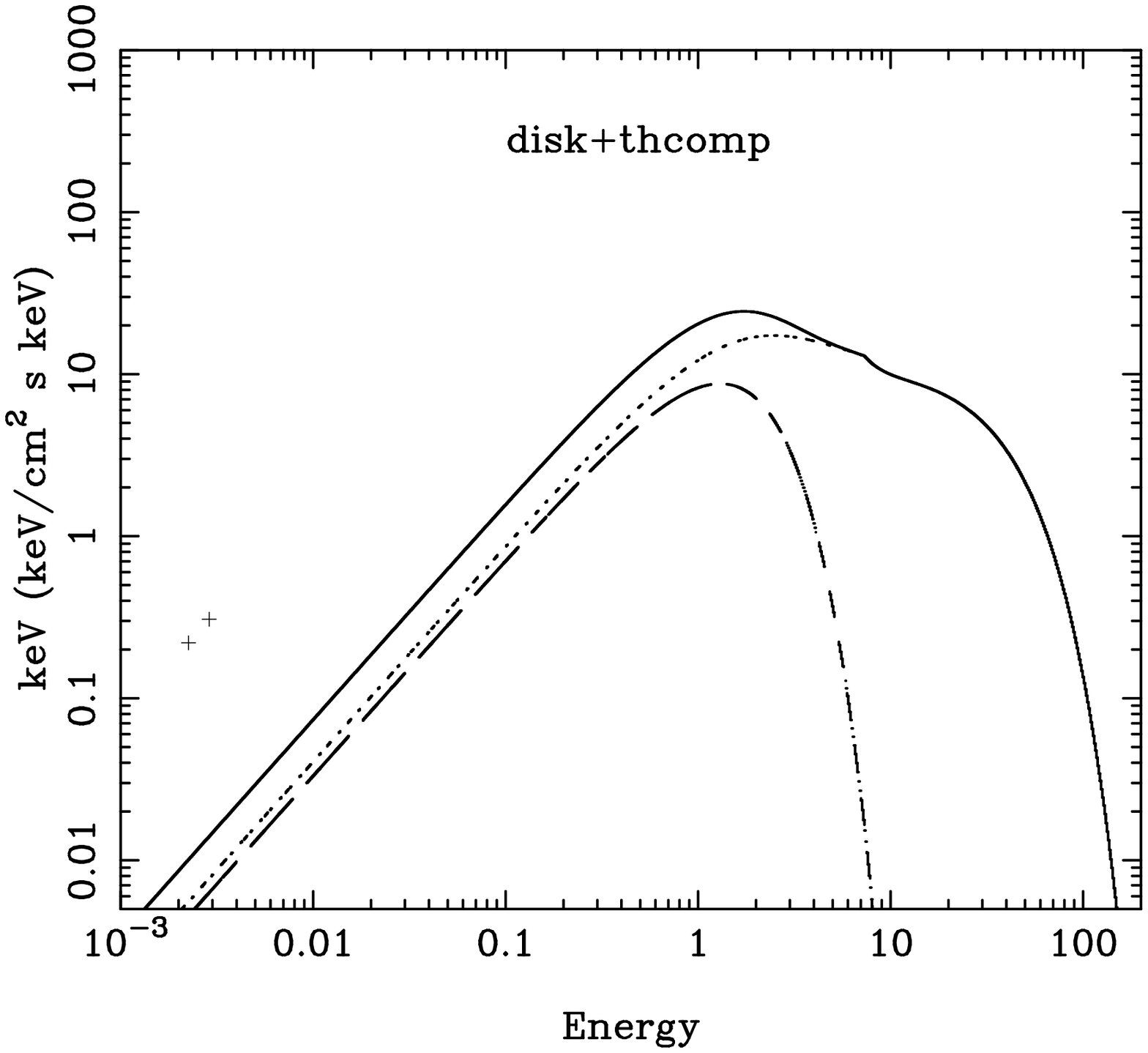}
\end{tabular}
\end{center}
\caption{Absorption corrected X-ray and optical spectra of the second
dataset. The left hand panel shows the best fit broken power law model for the
X-ray tail. The extrapolation of this down to the optical overpredicts the
simultaneously measured V and B data by a factor of over 1000. There must
be a low energy break in the spectrum. This is naturally produced in Comptonisation
models where the disc is the origin of the seed photons. The right hand
panel shows the best fit thermal Comptonisation model for the X-ray tail. 
This underestimates the optical flux, leaving room for a reprocessed flux
component from illumination of the disc and/or some nonthermal synchrotron
emission from the outer jet.}
\label{fig:opt_x}
\end{figure*}

The conflict with the broad iron line motivates us to review the evidence for 
Comptonisation as the emission mechanism for the tail seen in these data. The 
only feasible alternative appears to be the jet. X-ray emission from the
jet in the low/hard state has been proposed by e.g. Markoff et al (2001), so 
here we use this as a guide to what might be expected from these models in the
VHS. Direct synchrotron emission might give rise to a power law, or 
broken power law, or cutoff power law tail (Markoff et al 2001; Wilms et al 
2005; Markoff, Nowak \& Wilms 2005). In \S3 we showed that a power law gave a 
completely unacceptable fit to these data, but a broken power law is as good a 
description of the continuum as our thermal+nonthermal Comptonisation model 
(see Appendix). 
However, such models run into difficulties when extrapolated to lower energies 
as the steep spectrum dominates the bolometric luminosity. Fig. \ref{fig:opt_x}a 
shows this for the best fit broken power law and disc continuum model for the 
second dataset. The extrapolated X-ray continuum overpredicts the simultaneously 
observed optical V and B points (dereddedned using $A_v=4.75$) of Jain et al 
(1999) by a factor of over 1000. 

Clearly the data strongly require that there is a low energy break in the 
continuum between the soft X-ray and optical wavebands. Such a break can be 
produced from jet models at the point where the jet synchrotron emission becomes 
optically thick to self-absorption (e.g. Markoff, Nowak \& Wilms 2005; Giannios 
2005). However, there are some constrains on the energy of this break 
from variability arguments. The short timescale r.m.s. variability 
spectra of these VHS PCA data show a monotonic increase in fractional 
variability amplitude with energy form 3--50~keV, consistent with pivoting about 
a fixed point at $\sim 2$~keV (Gierli{\'n}ski \& Zdziarski 2005). There is no
obvious reason for jet models to pick out such an energy. 
This contrasts with models in which the X-ray continuum is produced by Compton 
scattering of disc photons. This naturally produces a break in the Comptonised 
continuum at energies below the seed photon energy (see Fig.~\ref{fig:opt_x}b), 
and since this is at $1-2$~keV there is no fine tuning required to explain this 
energy being picked out as the spectral pivot. 

Gierli{\'n}ski \& Zdziarski 
(2005) show that for the second dataset used here, the r.m.s. variability 
spectrum from the PCA is well matched by  thermal Compton scattering of the disc 
photons (see also Rodrigues et al 2004; Zdziarski et al 2005; Sobolewska \& 
$\dot{\rm Z}$ycki 2006 for the same results for other VHS spectra). The 
Comptonisation models actually underpredict the optical flux, but this is 
expected as the disc model here only includes direct emission, while 
reprocessing is likely to be an important contributor to the optical emission 
(e.g. Esin et al 2000; Hynes 2005), and optical synchrotron emission from the 
jet may also be present (see Turler et al 2004 for a model of the plateau/state 
C/class $\chi$ state of GRS 1915+105 with a turndown in the optical).

Given the strength of the disc emission in the VHS then it is also rather 
natural to assume that this is the dominant contributor to the seed photons. 
Again this contrasts with the low/hard state, where the disc is much weaker, and 
the jet might be expected to be a more important contributor (Markoff et al 
2001). However, even for the low/hard state, the jet models fail to reproduce 
the detailed shape of the tail (Zdziarski et al 2003), so that the current 
version of these jet models now have the majority of the X-ray emission produced 
by Comptonisation of the disc and cyclo--synchrotron photons by thermal 
electrons at the jet base (Markoff et al 2005). Since disc Comptonisation 
appears to be a component of the X-ray continuum even in the low/hard state, this 
process should be even more dominant with the stronger disc emission seen in the 
VHS. 

Thus both the multiwavelength spectral shape and energy dependence of the X-ray 
variability strongly support a disc Comptonisation interpretation of the X-ray 
tail. We show in \S7 that this is not inconsistent with the presence of high (or 
low) frequency QPO's. A hot (and so presumably geometrically thick) inner disc 
corona may actually be {\rm required} to produce the strong jet (McKinney 2005) 
traced by radio emission during this outburst of XTE~J$1550-564$ (Corbel et al 
2001). The only obvious inconsistency is with the interpretation of the iron 
line in these data given by Miller et al (2004). However, unlike the 
Comptonisation interpretation of the tail, which seems to be the only viable 
mechanism, there does exist an alternative explanation for the line shape. Done 
\& Gierli{\'n}ski (2005) show that the ASCA spectrum in the second VHS 
observation used here can be fit equally well by much less extreme line smearing 
if there is also some ionized absorption present. The underlying iron line width 
is then reduced,  consistent with originating from reflection in the outer disc, 
which allows the inner disc to be mostly covered by the corona.

Thus there are no data which unambiguously rule out an optically thick inner 
disc corona, while there are many observations which strongly support its 
existence. Such a corona will inevitably distort the disc spectrum in the ways 
explored in this paper. 

\section{Conclusions}

We investigate the broadband continuum spectra of two VHS observations
covering  0.8--200~keV. These have a strong X-ray tail, contributing more than
half of the total bolometric luminosity of the source. The disc in these spectra
has an apparent temperature which is much lower than 
that seen from the same source at the same luminosity in the disc dominated 
high/soft state. We show that this is not simply produced by the
effect of an optically thick corona hiding the emission from the inner disc,
but instead requires that the disc changes in some way from that seen in the
high/soft state, either in area (Fig.~\ref{fig:sketch2}a), 
emissivity (Fig.~\ref{fig:sketch2}b) or both (Fig.~\ref{fig:sketch2}c). 

The most obvious change in these spectra is that there is now a corona. 
Energetically, this must be powered ultimately by the accretion flow, so there 
is less power available for the disc itself. We build a specific example of 
disc-corona coupling, where the corona over the inner disc drains power from the 
disc material under it, changing its emissivity (Svensson \& Zdziarski 1996). We 
show that this reduces the amount of disc truncation required, to a level where 
it is probably not significant given the model uncertainties e.g. if some 
fraction of the accreting mass goes through the corona as well as some fraction 
of the power, or if the geometry of the corona is more quasi-spherical than 
slab-like. However, we marginally favour the coupled disc-corona system with a 
slightly truncated disc (Fig.~\ref{fig:sketch2}c) as this can simultaneously 
explain the LF QPO in the relativistic disc precession model. However, since 
there is as yet no clear consensus on models of QPO formation, this cannot give 
unambiguous information on disc radii. Thus Fig.~\ref{fig:sketch2}b, where the 
disc area (though not its properties!) is unchanged from the disc dominated 
states is also consistent with the spectra, though this leaves the change in 
LF QPO as an 
open question. Conversely, a large truncation radius with continuous corona 
(Fig.~\ref{fig:sketch2}a) {\em is} ruled out by the {\em steep} Comptonised 
spectrum seen.

However, we stress that irrespective of the specific model, the strong
Comptonisation seen in these very high state data clearly requires
that the underlying disc properties are rather different to those of a
disc without a corona. Unsurprisingly, the data require that the disc
structure {\em changes} in response to whatever changing dissipation
mode powers the growth of the coronal structure. 

%We review other constraints on the very high state spectra, and show
%that an optically thick Comptonisation in an inner disc corona is not
%inconsistent with the presence of high frequency QPO's.  However, it
%does conflict with claims of a strong, extremely broad iron line in
%these spectra. Either the iron line width is overestimated (perhaps due
%to the unrecognized presence of narrow absorption lines) or the tail is
%not formed by Comptonisation.

We also stress that these VHS spectra are very similar to those seen
from several ULX in terms of having a very low apparent disc
temperature for their luminosity.  This has been interpreted as
evidence for intermediate mass black holes in the ULX, of order $1000
M_\odot$ (Miller, Fabian \& Miller 2004). Obviously this is not an
option in XTE~J$1550-564$, where the black hole mass is constrained by
dynamical studies to be a normal stellar remnant black hole mass of
order $\sim 10 M_\odot$ (Orosz et al 2002).  We suggest that the same
process, of a corona draining energy from the inner disc, may be
operating in these ULX. Many of these would then be compatible with
fairly normal binary systems, where the black hole mass is at the
upper end of that which can reasonably form from models of stellar
collapse (typically $30 M_\odot$ or up to $\sim80~M_\odot$ from recent
calculations of binary evolution; Belczynski et al.  2004) accreting
at high mass accretion rates (0.1--few times the Eddington limit). We
will explore this possibility in depth in a subsequent paper.

\section{Acknowledgements}

The present work is supported in part by JSPS grant of No.13304014 and
No.14079101. CD thanks ISAS and RIKEN for their hospitality during
visits at which this work was carried out.
%by JSPS grant of No.13304014 (ISAS)
AK is supported by special postdoctoral researchers program in RIKEN.

\appendix
\section{Details of spectral fits}

We use the second dataset to quantify some aspects of the fits, with a baseline 
model of a broken power law tail (together with disc and SEG reflection gives 
$\chi^2_\nu=146.3/193$ for $kT_{disc}=0.67^{+0.04}_{-0.05}$~keV. By contrast, 
using the newest response ({\sc pcarsp} 10.1) gives $\chi^2_\nu=183.3/193$ for 
$kT_{disc}=0.77\pm 0.06$~keV. The shift is within the uncertainties, but more 
important is systematic: representative high/soft state spectra also have 
similarly shifted temperatures so that the difference between the disc 
properties in the VHS and high/soft state is robust. Replacing the
HEXTE data with a file without the 0.5 per cent systematic error gives
$\chi^2_\nu=147.6/193$ and $kT_{disc}=0.66^{+0.05}_{-0.04}$~keV, nearly
identical as the statistical uncertainties are all larger than 1 per cent. 

\begin{table*}
\centering
\begin{minipage}{175mm}
\caption{Best fit parameters obtained by a simultaneous ASCA/{\it
RXTE} observation.  The continuum is always described by a
multi-colour disc spectrum, plus a low temperature thermal comptonised
component, together with a power law extending to high energies. The
five different fits show the results of the different descriptions of
reflection. These are a narrow gaussian line and smeared edge in SEG,
while the other models include the full reflection spectrum (line and
continuum) in the {\it thcomp} model. In the case of Xi1, Xi2 and
Xi3, the data in 5--12~keV is excluded from the spectral fit, $R_{\rm
in}$ is fixed to 30~$R_{\rm g}$, and the value of $\xi$ is fixed to
$10^2$, $10^3$, $10^4$, for models Xi1, Xi2, and Xi3,
respectively. Normalization factors with the PCA data are used to
calculate luminosities.
}
\label{tab:asca-rxte}
\begin{tabular}{cccccccccc}
\hline \hline
model&$N_{\rm H}$ & $T_{\rm in}$ & $\Gamma_{\rm thc}$ & $T_{\rm e}$ & $L_{\rm disk}$ & $L_{\rm thc}$(slab) & $L_{\rm pow}$&   smedge,line (a) & $\chi^2/{\rm dof}$\\
&$10^{21}{\rm cm^{-1}}$ & keV & & keV &($N_{\rm disk}$)  &($N_{\rm thc}$) &  & reflection (b--e) & \\
\hline \hline
\multicolumn{10}{c}{Before peak}\\
\hline \hline
SEG&$6.5\pm0.2$ & $0.45^{+0.02}_{-0.01}$&$2.08^{+0.01}_{-0.02}$ &  $10.0\pm0.5$  & 1.03 &3.14&0.86 &   $E=7.9^{+0.3}_{-0.2}$ keV   &201.4/192\\
&& & && (37.6) &(58.0) &    &$\tau_{\rm max}=0.3^{+0.2}_{-0.1}$ &  \\
&& & && & & &   width$=3^{+2}_{-1}$~keV    &\\
&& & && & & &     $E=6.48^{+0.03}_{-0.07}$ keV &\\
&& & && & & &    EW=$81\pm10$ eV  &\\ 
 \hline
REF&$5.9\pm0.1$ &$0.51^{+0.01}_{-0.02}$ &$2.10\pm0.01$ 
&$11.6^{+0.5}_{-0.6}$
&$0.92$  &3.05/2.64 %\footnote{including <the reflection component}
& 0.76   & $\Omega/2\pi=0.23^{+0.04}_{-0.01}$    &230.2/194  \\
& & & &&($30.1$) &(51.0/44.9) &    &   $\xi=6.8^{+0.8}_{-2.7}\times 10^3$  &  \\
& & & && & &    &    $R_{\rm in}=94^{+159}_{-83}~R_{\rm g}$  & \\ \hline
Xi1&$6.5^{+0.2}_{-0.3}$  &$0.45\pm0.03$ &$2.27^{+0.02}_{-0.03}$ &$14\pm1$ &$0.86 $  & 3.52/2.95 &0.76    & $\Omega/2\pi=0.76^{+0.08}_{-0.06}$  &146.3/146   \\
& & & &&(31.0)  &(80.2/64.4) &    &   ($\xi=10^2$) &  \\ \hline
Xi2&$6.1^{+0.2}_{-0.1}$&$0.50\pm0.02$ &$2.14\pm0.01$ &$11.5\pm6$ &$0.90$  &3.14/2.69 &0.81    & $\Omega/2\pi=0.37\pm0.07$   &161.2/146   \\
& & & &&(29.8)  &(57.0/48.5) &    &   ($\xi=10^3$) & \\ \hline
Xi3&$6.2\pm0.2$&$0.48\pm0.02$ &$2.10\pm0.01$ &$11.3\pm0.6$ &0.97& 3.10/2.41 & 0.81  & $\Omega/2\pi=0.42^{+0.08}_{-0.04}$ &  157.8/146 \\
& & & &&(33.5)  &(53.5/42.7) &    &   ($\xi=10^4$) &  \\ 
\hline \hline
\multicolumn{10}{c}{After peak}\\
\hline \hline
SEG&$6.7\pm0.2$ & $0.55\pm0.02$&$2.32^{+0.02}_{-0.03}$ &  $12\pm1$  & 1.19 &4.09 &0.69 &   $E=7.5\pm0.2$ keV   &157.8/192\\
&& & && (36.1) &(80.9) &    &$\tau_{\rm max}=0.6\pm0.2$ &  \\
&& & && & & &   width$=5^{+3}_{-2}$~keV    &\\
&& & && & & &     $E=6.4\pm0.1$ keV &\\
&& & && & & &    EW=$41^{+12}_{-6}$ eV  &\\ 
 \hline
REF&$7.1\pm0.1$ &$0.44^{+0.06}_{-0.02}$ &$2.74\pm0.03$ 
&$>120$ \footnote{The best fit value appeared in $>200$~keV.}
&$0$  &5.87/5.32 %\footnote{including the reflection component}
& 0.29   & $\Omega/2\pi=1.7^{+0.2}_{-0.1}$    &226.2/194  \\
& & & &&($0$) &(163.9/161.9) &    &   $\xi=5^{+2}_{-1}$  &  \\
& & & && & &    &    $R_{\rm in}=12\pm2~R_{\rm g}$  & \\ \hline
Xi1&$7.1^{+0.3}_{-0.2}$  &$0.53\pm0.03$ &$2.64\pm 0.03$ &$38^{+18}_{-9}$ &$0.89 $  & 5.20/4.00 &0.46    & $\Omega/2\pi=1.6^{+0.3}_{-0.2}$  &113.5/146   \\
& & & &&(27.6)  &(139.5/98.5) &    &   ($\xi=10^2$) &  \\ \hline
Xi2&$6.5\pm0.3$&$0.60\pm0.02$ &$2.42\pm0.02$ &$16\pm2$ &$1.17 $  &4.10/3.29 &0.62    & $\Omega/2\pi=0.73^{+0.05}_{-0.09}$   &131.6/146   \\
& & & &&(32.9)  &(80.9/64.4) &    &   ($\xi=10^3$) & \\ \hline
Xi3&$6.6^{+0.3}_{-0.1}$&$0.55^{+0.02}_{-0.03}$ &$2.35^{+0.02}_{-0.01}$ &$14^{+2}_{-1}$ &1.13  & 4.18/3.16 & 0.64  & $\Omega/2\pi=0.62^{+0.04}_{-0.08}$ &  127.9/146   \\
& & & &&(34.2)  &(80.8/63.1) &    &   ($\xi=10^4$) &  \\ 
\hline
 \end{tabular}
\end{minipage}
\end{table*}

\begin{table*}
\centering
\begin{minipage}{175mm}
\caption{Same as Table~\ref{tab:asca-rxte} but with 
{\sc dkbbth} instead of {\sc diskbb} plus {\sc thcomp} to describe a
simple (uncoupled energetics) inner disc-corona.}
\label{tab:dkbbth}

\begin{tabular}{cccccccccc}
\hline \hline
model&$N_{\rm H}$ & $T_{\rm in}$ & $r_{\rm in}$ 
&$R_{\rm c}/r_{\rm in}$&$\Gamma_{\rm thc}$ & $T_{\rm e}$ & disk-corona\footnote{The unabsorbed bolometric luminosity of the {\sc dkbbth}, $L$, is calculated for inclination $70^\circ$ 
by integrating the model spectrum by 0.001--300~keV.  
The intrinsic
disc luminosity, $L_{\rm disk}^{\rm int}$, is calculated by referring to observed $r_{\rm in}$ and $T_{\rm in}$ as $4\pi r_{\rm in}^2 \sigma T_{\rm in}^4$.} 
& smedge& $\chi^2/{\rm dof}$\\
&$10^{21}{\rm cm^{-1}}$ & keV & [km] &  & &keV   &&reflection &\\
\hline \hline
\multicolumn{10}{c}{Before peak}\\
\hline \hline
SEG&$6.6^{+0.2}_{-0.3}$ & $ 0.60^{+0.04}_{-0.02}$&$129^{+10}_{-18}$& $ 1.59^{+0.03}_{-0.02}$ &$2.08^{+0.01}_{-0.02}$  &  $9.9^{+0.4}_{-0.6}$  &$L=4.20$%(4.17)
  & $E=8.0^{+0.2}_{-0.1}$ &204.9/192\\ 
   &&&&&&&$L_{\rm disc}^{\rm int}=2.79$ 
   &$\tau_{\rm max}=0.29^{+0.26}_{-0.08}$  &\\
      &&&&&&&$\tau=2.3$&${\rm width}=2.6^{+0.2}_{-1.6}$  &\\
      &&&&&&&&$E=6.48\pm0.07$  &\\
            &&&&&&&&${\rm EW}=80\pm10$ [eV] &\\
\hline
REF& $6.1 \pm0.2$&$0.64 \pm0.02$&$107 ^{+5}_{-6}$&$1.58 ^{+0.01}_{-0.02}$&$2.11\pm0.02$&$11.8 \pm0.4$&$L=4.04$ & $\Omega/2\pi=0.32 ^{+0.02}_{-0.05}$   &227.3/194\\
            &&&&&&&$L_{\rm disc}^{\rm int}=2.48$ &$\xi=(3.4 ^{+1.5}_{-1.2})\times10^3$  &\\
            &&&&&&&$\tau=2.0$&$R_{\rm in}=180 ^{+410}_{-100}$  &\\
\hline
Xi1 & $6.7\pm0.2$&$0.57^{+0.06}_{-0.04}$&$170^{+20}_{-50}$&$1.72^{+0.05}_{-0.04}$&$2.27\pm0.03$&$14\pm1$&$L=4.32$&$\Omega/2\pi=1.0\pm0.2$&148.9/146\\
                &&&&&&&$L_{\rm disc}^{\rm int}=3.81$&  &\\
                &&&&&&&$\tau=1.6$&  &\\
\hline
Xi2 & $6.2^{+0.3}_{-0.1}$&$0.63^{+0.05}_{-0.02}$&$110^{+20}_{-15}$&$1.57\pm0.03$&$2.12\pm0.01$&$11.4^{+0.5}_{-0.4}$&$L=4.04$&$\Omega/2\pi=0.48^{+0.10}_{-0.08}$&159.6/146\\
                &&&&&&&$L_{\rm disc}^{\rm int}=2.46$&  &\\
                &&&&&&&$\tau=2.0$&  &\\
\hline
Xi3 & $6.1^{+0.2}_{-0.4}$&$0.62^{+0.02}_{-0.04}$&$110^{+20}_{-7}$&$1.46\pm0.03$&$2.094^{+0.007}_{-0.008}$&$11.3^{+0.6}_{-0.5}$&$L=3.98$&$\Omega/2\pi=0.7\pm0.1$&157.0/146\\
                &&&&&&&$L_{\rm disc}^{\rm int}=2.31$&  &\\
                &&&&&&&$\tau=2.1$&  &\\
 \hline\hline
\multicolumn{10}{c}{After peak}\\
\hline \hline
SEG&$6.9\pm0.2$ & $0.71\pm0.03$&$110^{+6}_{-12}$ & $1.70\pm0.05$ &$2.32^{+0.02}_{-0.03}$  &  $12\pm1$  & $L=5.37$ & $E=7.6\pm0.2$  &151.1/192\\
     &&&&&&&$L_{\rm disk}^{\rm int}=3.97$&$\tau_{\rm max}=0.5^{+0.3}_{-0.2}$&\\
      &&&&&&&$\tau=1.7$&${\rm width}=4^{+3}_{-2}$  &\\
            &&&&&&&&$E=6.39\pm0.08$   &\\
            &&&&&&&& ${\rm EW}=50\pm10$[eV]  &\\
 \hline
REF&$6.6\pm0.1$&$0.72 ^{+0.02}_{-0.05}$&$105^{+11}_{-9}$&$1.9^{+0.2}_{-0.1}$&$2.46^{+0.08}_{-0.03}$&$20\pm1$&$L=5.33$ 	& $\Omega/2\pi= 0.64^{+0.13}_{-0.04}$  	  &228.5/194\\
            &&&&&&&$L_{\rm disc}^{\rm int}=3.83$ &$\xi=(1.4 ^{+0.3}_{-0.5})\times10^2$  &\\
            &&&&&&&$\tau=1.1 $&$R_{\rm in}=31 ^{+8}_{-9}$  &\\
\hline
Xi1 & $6.9\pm0.1$&$0.64^{+0.04}_{-0.03}$&$130\pm15$&$2.2^{+0.3}_{-0.2}$&$2.58^{+0.02}_{-0.03}$&$20^{+2}_{-1}$&$L=5.48$&$\Omega/2\pi=1.2^{+0.2}_{-0.4}$&127.0/146\\
                        &&&&&&&$L_{\rm disc}^{\rm int}=3.66$&  &\\
                &&&&&&&$\tau=1.0$&  &\\
\hline
Xi12& $6.8^{+0.2}_{-0.3}$&$0.70^{+0.03}_{-0.05}$&$110^{+17}_{-10}$&$1.7\pm0.1$&$2.40^{+0.02}_{-0.01}$&$16^{+2}_{-1}$&$L=5.35$&$\Omega/2\pi=0.6\pm0.1$&121.2/146\\
                        &&&&&&&$L_{\rm disc}^{\rm int}=3.75 $&  &\\
                &&&&&&&$\tau=1.4$&  &\\
\hline
Xi13& $6.8\pm0.2$&$0.66\pm0.03$&$120\pm10$&$1.52^{+0.03}_{-0.05}$&$2.355^{0.008}_{-0.012}$&$16\pm1$&$L=5.37$&$\Omega/2\pi=0.9^{+0.2}_{-0.1}$&124.2/146\\
                        &&&&&&&$L_{\rm disc}^{\rm int}=3.47$&  &\\
                &&&&&&&$\tau=1.4$&  &\\
\hline
 \end{tabular}
\end{minipage}
\end{table*}

\begin{table*}
\centering
\begin{minipage}{175mm}
\caption{Same as Table~\ref{tab:asca-rxte} but with {\sc dkbbfth} instead of 
{\sc diskbb} plus {\sc thcomp} to describe an inner disc-corona with
coupled energetics.}

\label{tab:dkbbfth}

\begin{tabular}{cccccccccc}
\hline \hline
model&$N_{\rm H}$ & $T_{\rm in}^{\rm int}$ & $r_{\rm in}$&$R_{\rm c}/r_{\rm in}$&$\Gamma_{\rm thc}$ & $T_{\rm e}$ & disc-corona\footnote{The unabsorbed bolometric luminosity of the {\sc dkbbfth}, $L$, is calculated for inclination $70^\circ$ 
by integrating the model spectrum by 0.001--300~keV.  
In this fit, the intrinsic disc luminosity calculated as 
$4\pi r_{\rm in}^2 \sigma {T_{\rm in}^{\rm int}}^4$}
&smedge & $\chi^2/{\rm dof}$\\
&$10^{21}{\rm cm^{-1}}$ & keV & [km] &  & keV &  &  &reflection & \\
\hline \hline
\multicolumn{10}{c}{Before peak }\\
\hline 
\hline
SEG&$6.4\pm0.1$ & $0.88^{+0.03}_{-0.04} $&$71^{+10}_{-5}$ & $ 2.46^{+0.05}_{-0.09}$ &$2.07\pm0.02$  &  $9.8\pm0.5$  & $L=4.03 $  &$E=8.0^{+0.2}_{-0.3}$
&204.7/192\\
&&&&&&&$f=0.58$&$\tau_{\rm max}=0.3^{+0.3}_{-0.1}$ & \\
&&&&&&&$\tau=2.3$  &${\rm width}=3^{+3}_{-1}$&\\
&&&&&&& &$E=6.48\pm0.07$&\\
&&&&&&&&${\rm EW}=85\pm10$&\\
\hline
REF&$6.1\pm0.1$&$0.91\pm0.02$&$65^{+3}_{-1}$&$2.3^{+0.07}_{-0.05}$&$ 2.111^{+0.007} _{-0.005}$&$11.6 ^{+0.8}_{-0.4}$&$L=3.94$ &$\Omega/2\pi= 0.31^{+0.04}_{-0.2}$ &230.9/194\\
&&&&&&&$f=0.56$& $\xi=(2.7 ^{+2.0}_{-0.5})\times10^3$ &\\
&&&&&&&$\tau=2.0$& $R_{\rm in}=160^{+>900}_{-80}$ &\\
\hline\hline

Xi1&$6.4\pm0.1$&$0.80^{+0.06}_{-0.04}$&$84^{+11}_{-13}$&$2.43^{+0.06}_{-0.07}$&$2.26^{+0.02}_{-0.03}$&$13.7 ^{+0.5}_{-1.1}$&$L=3.73$ &$\Omega/2\pi= 0.9\pm0.02$ &150.3/146\\
&&&&&&&$f=0.51$& &\\
&&&&&&&$\tau=1.7$&  &\\ \hline
%dkbbfth_c_after2.xcm after3.log

Xi2&$6.1\pm0.1$&$0.89^{+0.02}_{-0.05}$&$67^{+8}_{-3}$&$2.3^{+0.2}_{-0.1}$&$2.137^{+0.007}_{-0.009}$&$11.5 ^{+0.4}_{-0.8}$&$L=3.64$ &$\Omega/2\pi= 0.49^{+0.09}_{-0.10}$ &159.3/146\\
&&&&&&&$f=0.55$& &\\
&&&&&&&$\tau=2.0$&  &\\
\hline
Xi3&$6.1\pm0.1$&$0.83^{+0.04}_{-0.03}$&$75^{+2}_{-6}$&$2.11^{+0.06}_{-0.10}$&$2.101^{+0.002}_{-0.014}$&$11.6 ^{+0.2}_{-0.8}$&$L=3.43$ &$\Omega/2\pi= 0.70^{+0.16}_{-0.14}$ &157.9/146\\
&&&&&&&$f=0.56$& &\\
&&&&&&&$\tau=2.1$&  &\\

\hline\hline
\multicolumn{10}{c}{After peak}\\
\hline \hline
SEG&$ 6.7\pm0.2$ & $0.95^{+0.03}_{-0.02}$&$71 ^{+7 }_{-5}$ & $2.32^{+0.06}_{-0.07}$ &$2.31\pm0.02$  &  $12\pm1$  & $L=5.23$  & $E=7.6\pm0.2$&154.8/192\\
&&&&&&&$f=0.47$&$\tau_{\rm max}=0.5^{+0.3}_{-0.2}$ & \\
&&&&&&&$\tau=1.7$  &${\rm width}=4^{+1}_{-2}$&\\
&&&&&&& &$E=6.39\pm0.08$&\\
&&&&&&&&${\rm EW}=50\pm10$ &\\
 \hline
REF&$6.8^{+0.3}_{-0.1}$&$0.86^{+0.02}_{-0.09}$&$86^{+11}_{-7}$&$2.76^{+0.3}_{-0.2}$&$ 2.48^{+0.03} _{-0.02}$&$21.8 ^{+0.5}_{-1.7}$&$L=5.48$ &$\Omega/2\pi= 0.7\pm0.1$ &207.2/194\\
&&&&&&&$f=0.39$& $\xi=(1.1\pm0.2)\times10^2$ &\\
&&&&&&&$\tau=1.1$& $R_{\rm in}=26^{+9}_{-44}$ &\\

\hline
Xi1&$7.2^{+0.2}_{-0.1}$&$0.74^{+0.03}_{-0.04}$&$120\pm20$&$2.9\pm0.3$&$2.57^{+0.03}_{-0.05}$&$22^{+1}_{-2}$&$L=5.30$ &$\Omega/2\pi= 1.1^{+0.1}_{-0.2}$ &114.8/146\\
&&&&&&&$f=0.36$& &\\
&&&&&&&$\tau=0.97$&  &\\

\hline
Xi2&$6.7^{+0.2}_{-0.1}$&$0.89^{+0.03}_{-0.04}$&$78^{+2}_{-6}$&$2.20^{+0.04}_{-0.09}$&$2.41\pm0.01$&$16\pm1$&$L=4.93$ &$\Omega/2\pi= 0.62^{+0.10}_{-0.08}$ &123.2/146\\
&&&&&&&$f=0.43$& &\\
&&&&&&&$\tau=1.4$&  &\\
\hline
Xi3&$6.7^{+0.2}_{-0.1}$&$0.83\pm0.03$&$85^{+4}_{-6}$&$1.97^{+0.06}_{-0.10}$&$2.36^{+0.03}_{-0.01}$&$17\pm1$&$L=4.39$ &$\Omega/2\pi= 0.9^{+0.3}_{-0.1}$ &128.1/146\\
&&&&&&&$f=0.45$& &\\
&&&&&&&$\tau=1.4$&  &\\ 
%dkbbfth_e_after2.xcm after2.log
\hline
 \end{tabular}
\end{minipage}
\end{table*}


\begin{thebibliography}{99}
\bibitem{} Abramowicz M. A., Kluzniak W, 2001, A\& A, 374, 19
\bibitem{} Agol E., Krolik J. H., 2000, ApJ, 528, 161
%\bibitem{} Ballantyne D. R.,  Iwasawa K., Ross R. R., , MNRAS, 2002, MNRAS, 323, 506
\bibitem{} Ballantyne D. R.,  Ross R. R., Fabian A. C., MNRAS, 2001,
  MNRAS, 327, 10
\bibitem{} Blaes O., Arras P., Fragile C., 2006, MNRAS, submitted, 
(astro-ph/0601379)
%\bibitem{} Belloni T. et al , 1996, ApJ, 472, 107
%\bibitem{} Beloborodov A. M., 1998, MNRAS, 297, 739
\bibitem[\protect\citeauthoryear{Belczynski, Sadowski, \& 
Rasio}{2004}]{2004ApJ...611.1068B} Belczynski K., Sadowski A., Rasio F.~A., 
2004, ApJ, 611, 1068 
%\bibitem{} Churazov E., Gilfanov M., Revnivtsev M., 2001, MNRAS, 321, 759
\bibitem{} Corbel S., et al, ApJ, 554, 43
\bibitem[\protect\citeauthoryear{Coppi}{1999}]{1999ASPC..161..375C} Coppi 
P.~S., 1999, ASPC, 161, 375 
%\bibitem{} Cui W., Zhang S. N., Chen W., Morgan E. H., 1999, ApJ, 512, 43
\bibitem{} di Matteo T., Psaltis D., 1999, ApJ, 526, 101
\bibitem{} Davis S. W., Blaes O. M., Hubeny I., Turner N. J., 2005, ApJ, 621, 372
\bibitem{} Davis S. W., Done C., Blaes O. M., 2006, ApJ, in press
%\bibitem{} Done C., et al , 1992, ApJ, 395, 275
\bibitem{} Done C., Gierli\'nski M., 2003, MNRAS, 342, 1041
\bibitem{} Done C., Madejski G. M., $\dot{\rm Z}$ycki P. T., 2000, ApJ, 536, 213
\bibitem{} Done C., $\dot{\rm Z}$ycki P., Smith D. A., 2002, MNRAS, 331, 453 %cygx2
\bibitem[\protect\citeauthoryear{Done \& 
Gierli{\'n}ski}{2006}]{2006MNRAS.367..659D} Done C., Gierli{\'n}ski M., 
2006, MNRAS, 367, 659 
%\bibitem{} Ebisawa K., Mitsuda K., Hanawa T., 1991, ApJ, 367, 213
\bibitem[Ebisawa et al (1994)]{} Ebisawa K. et al, 1994, PASJ, 46, 375
%\bibitem{} Esin A., McClintock J. E., Narayan R., 1998, ApJ, 489, 865
%\bibitem{} Esin A., McClintock J., Narayan R. 1997, ApJ, 489, 865
\bibitem[\protect\citeauthoryear{Esin et al.}{2000}]{2000ApJ...532.1069E} 
Esin A.~A., Kuulkers E., McClintock J.~E., Narayan R., 2000, ApJ, 532, 1069 
\bibitem[\protect\citeauthoryear{Giannios}{2005}]{2005A&A...437.1007G} 
Giannios D., 2005, A\&A, 437, 1007 
\bibitem{} Gierli\'nski M., Done C. 2003, MNRAS, 342, 1083
\bibitem{} Gierli\'nski M., Done C. 2004, MNRAS, 347, 885 (GD04)
%\bibitem{} Gierli\'nski M., Zdziarski A. Z., et al , 1999, MNRAS, 309, 496
\bibitem[\protect\citeauthoryear{Gierli{\'n}ski \& 
Zdziarski}{2005}]{2005MNRAS.363.1349G} Gierli{\'n}ski M., Zdziarski A.~A., 
2005, MNRAS, 363, 1349 
\bibitem{} Gilfanov M., Revnivtsev M., Molkov S., 2003, A\& A, 410, 217
%\bibitem[Hannikainen et al. (2001)]{} Hannikainen D., Campbell-Wilson D., 
%Hunstead R., Mclntyre V., Lovell J., Reynolds J., Tzioumis T., 
%\& Wu, K. 2001, AP\&SS, 276, 45
\bibitem{} Haardt F., Maraschi L. 1993, ApJ, 413, 507
\bibitem[\protect\citeauthoryear{Hynes}{2005}]{2005ApJ...623.1026H} Hynes 
R.~I., 2005, ApJ, 623, 1026 
%\bibitem{} Hawley J. F., Balbus S. A., 2002, ApJ, 573, 736
%\bibitem{} Homan J. et al , 2001, ApJ, 132, 377
%\bibitem{} Kawaguchi T., 2003, ApJ, 593, 69
\bibitem{} Jain, R.K., Bailyn C.D., Orosz J.A., Remillard R.A., McClintock J.E., 1999, ApJL, 517, 131
\bibitem{} Kubota A., et al , 1998, PASJ, 50, 667
\bibitem{kd04} Kubota A., Done C., 2004, MNRAS, 353, 980 (KD04)
\bibitem{kme03} Kubota A., Makishima K.,  Ebisawa K., 2001, ApJ, 560, L147
\bibitem{km04} Kubota A., Makishima K., 2004, ApJ, 601, 428 
%\bibitem{}Magdziarz P., Zdziarski A., 1995, MNRAS, 273, 837
\bibitem[]{} Makishima K., et al  1996, PASJ, 48, 171
\bibitem[\protect\citeauthoryear{Markoff, Falcke, \& 
Fender}{2001}]{2001A&A...372L..25M} Markoff S., Falcke H., Fender R., 2001, 
A\&A, 372, L25 
\bibitem[\protect\citeauthoryear{Markoff, Nowak, \& 
Wilms}{2005}]{2005ApJ...635.1203M} Markoff S., Nowak M.~A., Wilms J., 2005, 
ApJ, 635, 1203 
\bibitem{}McClintock J. E., Remillard R. A. \ 2003,
in Compact Stellar X-ray Sources, eds. Lewin W. H. G.,  \& van der Klis, M.,
(Cambridge University Press, Cambridge), in press (astro-ph/0306213)
\bibitem{} McKinney J.C., 2005, ApJ, 630, 5  
%\bibitem{}Mendez R. H., van der Klis M., 1997, ApJ, 479, 926
\bibitem{} Merloni A., Fabian A. C., Ross R. R., 2000, MNRAS, 313, 193
\bibitem{} Miller J. M., Fabian A. C., Miller M. C., 2004, ApJ, 614, 117
\bibitem{} Miller J.M., et al. 2003, MNRAS, 338, 7
\bibitem{} Miller J.M., Fabian A.C., Nowak M.A., Lewin W.H.G., 2004, to
appear in proc. 10th Anual Marcel Grossmann Meeting on General Relativity
(astro-ph/0402101)
\bibitem[Mitsuda et al (1984)]{} Mitsuda K., et al  1984, PASJ, 36, 741
\bibitem[Miyamoto et al (1991)]{}Miyamoto S., Kimura K., Kitamoto S., Dotani
 T., Ebisawa K. 1991, ApJ,  383, 784
%\bibitem{} Morrison R., McCammon D., 1983, ApJ, 270, 119
%\bibitem{} Nayakshin S., Kazanas D., Kallman T. R., 2000, ApJ, 537, 833 %thermal instability photoionized reflection
%\bibitem{} Narayan, R., Yi, I. 1995, ApJ, 452, 710
\bibitem[Orosz et al (2002)]{} Orosz J. A. et al  \ 2002, ApJ, 568, 845 
\bibitem{} Pozdnyakov L. A., Sobol I. M., Sunyaev R. A. 1983, ASPR, 2, 189
\bibitem[Remillard et al (2002)]{} Remillard R. A., Sobczak G. J., 
Muno M. P., McClintock J. E., 2002, ApJ, 564, 962
\bibitem[\protect\citeauthoryear{Remillard}{2005}]{2005AN....326..804R} 
Remillard R.~A., 2005, AN, 326, 804 
\bibitem[\protect\citeauthoryear{Rodriguez et 
al.}{2004}]{2004ApJ...615..416R} Rodriguez J., Corbel S., Hannikainen 
D.~C., Belloni T., Paizis A., Vilhu O., 2004, ApJ, 615, 416 
%\bibitem{} Psaltis D., Belloni T., van der Klis M., 1999, ApJ, 520, 262
%\bibitem{} Revnivtsev, M., Gilfanov M., Sunyaev R., Jahoda K., Markwardt C. 2003, A\&A, 411, 329
%\bibitem{} Rothschild R. E. et al   \ 1998, ApJ, 496, 538
%\bibitem{} Ross R. R., Fabian A. C., Ballantyne D. R., 2002, MNRAS, 336, 315
\bibitem{} Ross R. R., Fabian A. C., Young A. J., 1999, 306, 461
%\bibitem{} Rybicki G. B., Lightman A. P., 1979, in Radiative process in astrophysics, 
%(John Wiley \& Sons, Inc)
\bibitem{ss73} Shakura N., Sunyaev R., 1973,  A\&A, 24, 337 
%\bibitem[Shimura \& Takahara (1995)]{} Shimura T., Takahara F.,1995, ApJ, 445, 780
%\bibitem[Smith 1998]{} Smith, D.A. 1998, IAUC 7008
\bibitem[\protect\citeauthoryear{{\.Z}ycki \& 
Sobolewska}{2005}]{2005MNRAS.364..891Z} {\.Z}ycki P.~T., Sobolewska M.~A., 
2005, MNRAS, 364, 891 
\bibitem[Sobczak et al (1999)]{} Sobczak G. J., McClintock J. E., Remillard R. A., Levine A. M., Morgan E. H,
Bailyn C. D., Orosz J. A. 1999, ApJ 517, 121
%\bibitem[Sobczak et al (2000a)]{} Sobczak G. J., McClintock J. E., 
%Remillard R. A., Cui W., Levine A. M., Morgan E. H., \& Bailyn C. D. 2000a, 
%ApJ, 531, 53
%\bibitem[Sobczak et al (2000b)]{} Sobczak G. J., McClintock J. E., 
%Remillard R. A., Cui W., Levine A. M., Morgan E. H., \& Bailyn C. D. 2000b, 
%ApJ, 544, 993
\bibitem{}Stella L., Mario V., Morsink S. M., 1999, ApJ. 524, L63
\bibitem{} Svensson R., Zdziarski A. A., 1994, 436, 599 
\bibitem{} Tanaka Y., \& Lewin W. H. G. 1995, in X-ray
Binaries, eds. Lewin W. H. G.,  van Paradijs J. and van den Heuvel W. P. J.,
             (Cambridge University Press, Cambridge), p126
%\bibitem{} Turner N. J., 2004, ApJ, 605, L45
\bibitem[\protect\citeauthoryear{T{\"u}rler et 
al.}{2004}]{2004A&A...415L..35T} T{\"u}rler M., Courvoisier T.~J.-L., Chaty 
S., Fuchs Y., 2004, A\&A, 415, L35 
%\bibitem{}van der Klis M.,  \ 2004,
%in Compact Stellar X-ray Sources, eds. Lewin W. H. G.,  \& van der Klis, M.,
%(Cambridge University Press, Cambridge), in press (astro-ph/0410551)
\bibitem{} van der Klis M., 2000, ARA\&A, 38, 717
\bibitem[\protect\citeauthoryear{Wilms et al.}{2006}]{2006A&A...447..245W} 
Wilms J., Nowak M.~A., Pottschmidt K., Pooley G.~G., Fritz S., 2006, A\&A, 
447, 245 
%\bibitem[Wilson \& Done (2001)]{} Wilson C. D., Done C. 2001, MNRAS, 325, 167
%\bibitem[Wilson et al (1998)]{} Wilson C. A., Harmon B. A., Paciesas W. S., McCollough M. L. 1998, IAUC 7010
%\bibitem[]{} Wu K. et al 2002, ApJ, 565, 1161
\bibitem{} Zdziarski A. A.,  Johnson W. N., Magdziarz P.,1996, MNRAS, 283, 193
\bibitem{} Zdziarski A. A., Grove J. E., Poutanen J., Rao A. R., Vadawale 
S. V., 2001, ApJ, 554, 45
\bibitem[\protect\citeauthoryear{Zdziarski et 
al.}{2005}]{2005MNRAS.360..825Z} Zdziarski A.~A., Gierli{\'n}ski M., Rao 
A.~R., Vadawale S.~V., Miko{\l}ajewska J., 2005, MNRAS, 360, 825 
\bibitem[\protect\citeauthoryear{Zdziarski et 
al.}{2003}]{2003MNRAS.342..355Z} Zdziarski A.~A., Lubi{\'n}ski P., Gilfanov 
M., Revnivtsev M., 2003, MNRAS, 342, 355 
%\bibitem{} $\dot{\rm Z}$ycki, P. T., Done, C. \& Smith, D. A., 1998, ApJ, 496, 25
\bibitem{} $\dot{\rm Z}$ycki, P. T., Done, C. \& Smith, D. A., 1999, MNRAS, 305, 231

\end{thebibliography}
\end{document}